\documentclass[11pt,letterpaper]{article}
\usepackage[margin=1in]{geometry}
\usepackage[utf8]{inputenc}
\usepackage[T1]{fontenc}
\usepackage{microtype}

\usepackage{graphicx}

\usepackage{amsfonts,amssymb}
\usepackage{amsthm}
\usepackage{csquotes}
\usepackage{mathtools}
\usepackage{xspace}
\usepackage{enumerate}
\usepackage[ruled]{algorithm2e}
\usepackage{paralist}
\usepackage{bm}
\usepackage{dsfont}
\usepackage{enumitem}
\usepackage{comment}
\usepackage{svg}
\usepackage{complexity}

\usepackage{cite}

\usepackage{url}
\usepackage[colorlinks]{hyperref} %
\hypersetup{hypertexnames=false,citecolor=green!60!black,linkcolor=blue!60!black,urlcolor=blue!60!black}

\usepackage[capitalize,noabbrev,nameinlink]{cleveref}
\newtheorem{theorem}{Theorem}
\newtheorem{lemma}[theorem]{Lemma}
\newtheorem{corollary}[theorem]{Corollary}
\newtheorem{proposition}[theorem]{Proposition}
\theoremstyle{definition}
\newtheorem{definition}[theorem]{Definition}

\usepackage{soul}

\usepackage{xcolor}
\definecolor{mymagenta}{RGB}{200,60,140}
\definecolor{myblue}{RGB}{50,100,200}
\definecolor{myred}{RGB}{210,60,60}
\definecolor{myorange}{RGB}{235,140,40}
\definecolor{mygreen}{RGB}{60,170,90}
\definecolor{mypurple}{RGB}{120,80,180}

\usepackage[small]{caption}
\usepackage{subcaption}
\usepackage{tikz}
\usepackage{pgfplots}
\pgfplotsset{compat=newest}

\usepackage{booktabs,tabularx}

\usepackage{etoolbox}

\allowdisplaybreaks

\newcommand{\eps}{\varepsilon}

\newcommand{\RR}{\mathbb{R}}

\renewcommand{\cS}{\mathcal{S}}

\newcommand{\OPT}{{\mathrm{OPT}}}

\usepackage[disable,colorinlistoftodos,prependcaption,textsize=scriptsize]{todonotes}

\begin{document}
\pagestyle{plain}
\title{The Power of Migration on Identical Machines is Two}

\author{Nicole Megow\thanks{Faculty of Mathematics and Computer Science, University of Bremen, Germany. Email: \texttt{\{nicole.megow,bart.zondervan\}@uni-bremen.de} This work was supported by supported by the Deutsche Forschungsgemeinschaft (DFG, German Research Foundation) grant no.\ 517912373 and under Germany’s Excellence Strategy – EXC 3036 The Martian Mindset – project number 533607631.} \and Bart Zondervan\footnotemark[1]}
\date{}

\maketitle 

\begin{abstract} 
    We study the power of migration in preemptive scheduling on identical parallel machines with time windows.
    If an instance admits a migratory schedule on~$m$ machines, how many machines are needed to schedule it without migration?
    Kalyanasundaram and Pruhs~(2001) showed that $6m-5$ machines suffice and gave a polynomial-time construction using $12m-5$ machines, whereas $2m-1$ machines may be necessary.
    We close this gap, which has been open for more than 25~years: $2m-1$ machines always suffice, and a corresponding non-migratory schedule can be computed in strongly polynomial time.

    Our proof has two main technical ingredients.
    First, we introduce supercanonical schedules and prove that the conflict graph of every such schedule on $m$ machines is $(2m-1)$-colorable.
    The key property is that whenever a job pauses, the jobs running alongside it just before and just after the pause separate the jobs processed entirely during the pause from the rest of the conflict graph.
    We use these separators and Kempe exchanges to compute a coloring in polynomial time.
    Second, we construct a supercanonical schedule with polynomially many processing blocks by shrinking job windows as far as possible while preserving feasibility and solving a min-cost flow problem.
\end{abstract}

\section{Introduction}

In preemptive scheduling on parallel machines, a natural source of flexibility is \emph{migration}: an interrupted job may resume on a different machine.
Migration can make it easier to meet deadlines, but in practice it incurs communication and synchronization overhead.
This raises a question: how much does migration actually help?

We study this question for scheduling jobs with time windows on identical parallel machines, in the offline setting where all jobs are known in advance.
Each job has a processing time and must be processed entirely within its time window, given by a release date and a deadline.
Jobs may be preempted, that is, their processing may be interrupted and resumed later, but no job may run on two machines simultaneously.
A schedule is migratory if a preempted job may resume processing on a different machine, whereas in a non-migratory schedule each job is processed entirely on one machine. 
The \emph{power of migration} measures how much the number of machines needed for a feasible schedule can increase when migration is disallowed. 
More precisely, for a given $m$, we ask for the smallest number of machines sufficient to schedule without migration any job set that admits a feasible migratory schedule on $m$ machines.

Kalyanasundaram and Pruhs~\cite{KalP01} showed that the power of migration is constant: every instance feasible with migration on $m$ machines admits a non-migratory schedule on $6m-5$ machines.
Their constructive proof yields a pseudopolynomial-time algorithm, and they also give a polynomial-time algorithm using $12m-5$ machines.
Conversely, a simple instance with identical release dates and deadlines shows that $2m-1$ machines may be necessary~\cite{KorenDA00}.

In this paper, we close this gap, which has been open for more than 25~years, by proving that~$2m-1$ machines always suffice and constructing such a non-migratory schedule in strongly polynomial time.

\subsection{Our Results and Techniques}
Our main result is a sharp worst-case bound on the number of machines needed to eliminate migration.

\begin{theorem}
    \label{thm:main}
    Every instance admitting a feasible migratory schedule on~$m$ identical machines admits a feasible non-migratory schedule on $2m-1$ identical machines. Such a schedule can be computed in strongly polynomial time.
\end{theorem}

The bound is tight for every $m$~\cite{KorenDA00}.
Consider $2m-1$ jobs with common window $[0,1)$ and processing time $m/(2m-1)$.
McNaughton's wrap-around rule~\cite{McNaughton1959} schedules them migratorily on $m$ machines.
Since each job requires more than half the window, a non-migratory schedule needs one machine per job.

We follow the conflict-graph approach of Kalyanasundaram and Pruhs~\cite{KalP01}.
The vertices of the conflict graph are the jobs, and two jobs are adjacent whenever they are processed simultaneously.
A proper coloring assigns each job to one machine and therefore gives a non-migratory realization of the schedule.
The task is to construct a feasible schedule on~$m$ machines whose conflict graph can be colored with $2m-1$ colors.

Kalyanasundaram and Pruhs~\cite{KalP01} use canonical schedules, in which jobs that complete earlier receive priority.
Their density argument guarantees a vertex of low degree in every induced subgraph, yielding a coloring bound via the Szekeres--Wilf theorem~\cite{SzekeresW68}.
We show that the conflict graph of every canonical schedule is $(3m-2)$-colorable and that this bound is tight~(\Cref{appendix:canonical-3m-2}). 
Thus, canonicity alone does not suffice for the desired coloring bound.

We strengthen canonicity by excluding a second type of swap, which delays the start of a job, and call the resulting schedules supercanonical.
Even for these schedules, the minimum degree of the conflict graph may exceed $2m-2$, e.g., for $m=2$ (\Cref{appendix:supercanonical_min_degree}).
Thus, this minimum-degree approach does not suffice for our target bound, and we need a different coloring argument.

We show that the conflict graph of every supercanonical schedule on~$m$ machines can be $(2m-1)$-colored in time polynomial in the number of jobs and processing blocks (\Cref{sec:coloring-supercanonical-conflicts}).
Our key structural insight concerns pauses: whenever a job pauses, the jobs running alongside it immediately before and after the pause separate jobs processed entirely during the pause from the remaining conflict graph.
We color inductively: we remove an earliest-completing job, color the rest, and reinsert the job via Kempe exchanges~\cite{Kempe1879} across pause separators.
Each exchange either frees a color in the neighborhood of the inserted job or decreases a potential based on the last blocks in which colors meet this job.

Our second ingredient is a strongly polynomial-time construction of a supercanonical schedule with $O(n^2)$ processing blocks on the original $m$ machines, where $n$ is the number of jobs (\Cref{sec:poly-time}).
We first shrink the job windows while preserving feasibility until no single window can be shrunk further.
This excludes delay swaps and fixes the order of completion times independently of the schedule.
We then optimize a min-cost flow objective that favors earlier processing of jobs that complete earlier.
Together with a greedy realization of the resulting allocation, this excludes priority swaps.

\paragraph{Consequences and extensions}
Our migration-elimination theorem also has consequences for other scheduling problems.
It determines the exact worst-case loss in optimal weighted throughput when migration is forbidden.
On the same $m$ machines, a non-migratory schedule can always retain at least an $m/(2m-1)$ fraction of the migratory optimum (\Cref{sec:throughput}).
This guarantee is tight even for weights proportional to processing times, as shown by an example of Koren, Dar, and Amir.
For feasibility, doubling the speed of the original $m$ machines suffices to eliminate migration, and this speedup is asymptotically optimal (\Cref{sec:speed-augmentation}).
Under restricted assignment, however, even two eligible machines per job can force an $\Omega(\log m)$ factor increase in the number of machines~(\Cref{sec:restricted-assignment}).

\subsection{Further Related Work}

The migration-elimination bound of~\cite{KalP01} is used to derive approximation and competitive guarantees for throughput maximization against a migratory optimum, both offline~\cite{KarakostasK25} and online~\cite{EberleMS23,Eberle24,MoseleyPSZ24}.

An alternative to additional machines is \emph{speed augmentation}.
Chan, Lam, and To~\cite{ChanLT05} give a non-migratory online algorithm on $m$ machines of speed $3+2\sqrt{2}$ for instances feasible migratorily on $m$ unit-speed machines, and show that speed $2m/(m+1)$ may be necessary.
Devanur and Kulkarni~\cite{DevanurK18} give a polynomial-time algorithm for non-migratory deadline scheduling on unrelated machines with speed augmentation~$16$.
On identical machines, the non-migratory feasibility condition of Baruah and Fisher~\cite{BaruahFisher_RTS2008_NonMigratory_Feasibility} implies a speedup bound of $4-1/m$; see~\Cref{sec:speed-augmentation} for the derivation and comparison with our bound.

In the online setting, migration can have a much stronger effect.
Chen et~al.~\cite{ChenMS16} show that no deterministic non-migratory online algorithm can bound its machine requirement solely by a function of the migratory offline optimum.

A related line of work studies the \emph{power of preemption}, measured by the ratio between optimal non-preemptive and preemptive makespans.
For jobs available at time zero, this ratio is tightly bounded by $2m/(m+1)$ on $m$ identical machines~\cite{BraunS03} and by~$4$ on unrelated machines~\cite{CorreaSV12}.

\section{Preliminaries and Coloring Tools}
\label{sec:preliminaries}

Let~$J$ be a set of~$n$ jobs such that each job~$j\in J$ has a processing time~$p_j\in \RR_{>0}$, a release date~$r_j\in \RR_{\geq0}$ and a deadline~$d_j\in \RR_{\geq0}$.
For algorithmic statements, we assume that all numerical input data are rational and encoded in binary.
Structural statements also hold for arbitrary real input data.
A \emph{schedule}~$\cS$ of jobs~$J$ consists of a sequence of breakpoints $b_1 < b_2 < \cdots < b_{T+1}$ and processing sets~$\cS_1,\dots, \cS_T \subseteq J$.
For each~$t\in [T] \coloneqq \{1,\dots,T\}$, precisely the jobs in~$\cS_t$ are processed throughout the block~$B_t \coloneqq [b_t,b_{t+1})$.
The schedule is feasible on~$m$ identical unit-speed machines if
\begin{enumerate}
    \item[(1)] $|\cS_t| \leq m$ for every~$t\in [T]$, 
    \item[(2)] $j\in \cS_t$ implies~$B_t \subseteq [r_j,d_j)$ for every~$j\in J$, $t\in [T]$, and 
    \item[(3)] $\sum_{t\in [T] : j\in \cS_t} (b_{t+1} - b_t) = p_j$ for every job~$j\in J$.
\end{enumerate}
A machine realization of~$\cS$ on~$m$ machines consists of an injective assignment~$\mu_t\colon \cS_t \to \{1,\dots,m\}$ for every~$t\in [T]$.
Such an assignment exists whenever~$|\cS_t|\leq m$.
A job may change machines between blocks, which is called migration.
The realization is \emph{non-migratory} if there exists a single assignment~$c\colon J\to \{1,\dots,m\}$ such that~$\mu_t(j)=c(j)$ for all~$t\in [T]$ and~$j\in \cS_t$.

\subsection{Canonical and Supercanonical Schedules}
Let~$\cS$ be a feasible schedule of~$J$ on~$m$ machines.
For each job~$j$, let $\rho_j$ and $\delta_j$ denote its first and last processing blocks, that is, $\rho_j \coloneqq \min\{t\in [T] \mid j\in \cS_t\}$ and $\delta_j \coloneqq \max\{t\in [T] \mid j\in\cS_t\}$.
Let~$\prec$ order the jobs by nondecreasing~$\delta_j$-value, with ties broken by job index.
Kalyanasundaram and Pruhs~\cite{KalP01} introduced canonical schedules, in which jobs that complete earlier receive priority.
\begin{definition} 
    \label{def:priority_swap}
    A \emph{priority swap} exists for jobs~$j_1\prec j_2$ and blocks~$t_1<t_2$ if $\rho_{j_1} \leq t_1$, $j_2 \in \cS_{t_1}\setminus\cS_{t_2}$, and $j_1 \in \cS_{t_2}\setminus\cS_{t_1}$.
    Schedule~$\cS$ is \emph{canonical} if it admits no priority swap.
    We then call~$\prec$ its \emph{canonical ordering}.
\end{definition}

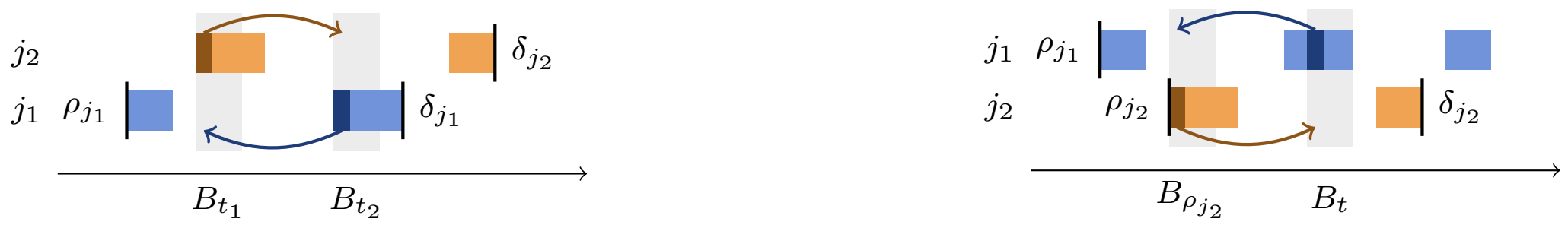
\begin{figure}[t]
    \centering
    \begin{subfigure}[b]{0.48\textwidth}
        \centering
        \begin{tikzpicture}[x=0.4cm,y=0.5cm,every node/.style={font=\scriptsize}]
            \fill[gray!15] (3,-0.2) rectangle (4,2.2);
            \fill[gray!15] (6,-0.2) rectangle (7,2.2);
            \fill[myblue!70] (1.5,0.15) rectangle (2.5,0.85);
            \fill[myblue!70] (6,0.15) rectangle (7.5,0.85);
            \fill[myblue!60!black] (6,0.15) rectangle (6.35,0.85);
            \draw[thick] (1.5,0) -- (1.5,1) node[midway,left] {$\rho_{j_1}$};
            \draw[thick] (7.5,0) -- (7.5,1) node[midway,right] {$\delta_{j_1}$};
            \fill[myorange!80] (3,1.15) rectangle (4.5,1.85);
            \fill[myorange!60!black] (3,1.15) rectangle (3.35,1.85);
            \fill[myorange!80] (8.5,1.15) rectangle (9.5,1.85);
            \draw[thick] (9.5,1) -- (9.5,2) node[midway,right] {$\delta_{j_2}$};
            \draw[->,thick,myorange!60!black] (3.175,1.85) to[bend left=25] (6.175,1.85);
            \draw[->,thick,myblue!60!black] (6.175,0.15) to[bend left=25] (3.175,0.15);
            \draw[->] (0,-0.6) -- (11.5,-0.6);
            \node at (3.5,-1.1) {$B_{t_1}$};
            \node at (6.5,-1.1) {$B_{t_2}$};
            \foreach \y/\lab in {0/j_1,1/j_2}
                \node[left] at (0,\y+0.5) {$\lab$};
        \end{tikzpicture}
    \end{subfigure}
    \hfill
    \begin{subfigure}[b]{0.48\textwidth}
        \centering
        \begin{tikzpicture}[x=0.4cm,y=0.5cm,every node/.style={font=\scriptsize}]
            \fill[gray!15] (3,-0.2) rectangle (4,2.2);
            \fill[gray!15] (6,-0.2) rectangle (7,2.2);
            \fill[myorange!80] (3,0.15) rectangle (4.5,0.85);
            \fill[myorange!60!black] (3,0.15) rectangle (3.35,0.85);
            \fill[myorange!80] (7.5,0.15) rectangle (8.5,0.85);
            \draw[thick] (3,0) -- (3,1) node[midway,left] {$\rho_{j_2}$};
            \draw[thick] (8.5,0) -- (8.5,1) node[midway,right] {$\delta_{j_2}$};
            \fill[myblue!70] (1.5,1.15) rectangle (2.5,1.85);
            \fill[myblue!70] (5.5,1.15) rectangle (7,1.85);
            \fill[myblue!60!black] (6,1.15) rectangle (6.35,1.85);
            \fill[myblue!70] (9,1.15) rectangle (10,1.85);
            \draw[thick] (1.5,1) -- (1.5,2) node[midway,left] {$\rho_{j_1}$};
            \draw[->,thick,myorange!60!black] (3.175,0.15) to[bend right=25] (6.175,0.15);
            \draw[->,thick,myblue!60!black] (6.175,1.85) to[bend right=25] (3.175,1.85);
            \draw[->] (0,-0.6) -- (11.5,-0.6);
            \node at (3.5,-1.1) {$B_{\rho_{j_2}}$};
            \node at (6.5,-1.1) {$B_t$};
            \foreach \y/\lab in {0/j_2,1/j_1}
                \node[left] at (0,\y+0.5) {$\lab$};
        \end{tikzpicture}
    \end{subfigure}
    \vspace*{-2ex}
    \caption{The configurations excluded in a supercanonical schedule.
    Time runs horizontally, and each job has a row, ordered by~$\prec$ from bottom to top, colored whenever the job is processed.
    Machine assignments are not shown.
    Left:~A priority swap for $j_1\prec j_2$ at blocks $t_1<t_2$.
    Right:~A delay swap for~$j_1$, $j_2$ at blocks $\rho_{j_2}<t$.}
    \label{fig:swaps}
\end{figure}

Suppose there exists a priority swap for jobs~$j_1\prec j_2$ at blocks~$t_1<t_2$, as in~\Cref{fig:swaps}.
For small~$\varepsilon>0$, there is a swap, where we process~$j_1$ instead of~$j_2$ during~$[b_{t_1},b_{t_1}+\varepsilon)$ and~$j_2$ instead of~$j_1$ during~$[b_{t_2},b_{t_2}+\varepsilon)$.
The result is again a feasible schedule (with different breakpoints).
We strengthen canonicity by excluding a second configuration, which postpones the start of a job.

\begin{definition} 
    \label{def:delay_swap}
    A \emph{delay swap} exists for distinct jobs~$j_1$ and $j_2$ at blocks~$\rho_{j_2} < t$ if $\rho_{j_1}<\rho_{j_2}<t<\delta_{j_2}$, $j_1\in \cS_t\setminus\cS_{\rho_{j_2}}$,  and~$j_2\in \cS_{\rho_{j_2}}\setminus \cS_t$.
    Schedule~$\cS$ is \emph{supercanonical} if it is canonical and admits no delay swap.
\end{definition}
Thus, in a supercanonical schedule, if~$j_1$ starts before~$j_2$ but is not processed in~$B_{\rho_{j_2}}$, then it can run in
a block~$B_t$ with~$t\in (\rho_{j_2},\delta_{j_2})$ only if~$j_2$ also runs in that block (\Cref{fig:swaps}).
Both canonicity and supercanonicity are preserved under deleting jobs while retaining the block partition.
The remaining jobs keep their first and last processing blocks and their relative order in~$\prec$, so any swap in the restricted schedule would already exist in the original schedule.
In~\Cref{sec:poly-time}, we construct a supercanonical schedule with~$O(n^2)$ blocks in polynomial time.

\subsection{The Conflict Graph}
\label{subsec:conflict_graph}

\begin{definition}
    \label{def:conflict_graph}
    The \emph{conflict graph} of a schedule~$\cS$ of jobs~$J$ is the simple undirected graph~$G^\cS = (J,E^\cS)$, where two distinct jobs~$j,k\in J$ are adjacent if there exists~$t\in [T]$ such that~$j,k\in \cS_t$, that is, both are processed in block~$t$.
\end{definition}
A \emph{$q$-coloring} of a graph~$H=(V,E)$ is a map $c\colon V\to[q]$ such that~$c(u)\neq c(v)$ for every edge~$\{u,v\}\in E$. 
The smallest such~$q$ is the \emph{chromatic number}~$\chi(H)$. 
Assigning jobs of each color to one machine gives a non-migratory realization, since jobs of the same color never run simultaneously.
Conversely, each non-migratory realization defines a coloring.
Thus, $\chi(G^\cS)$ is the minimum number of machines needed for a non-migratory realization of~$\cS$.

For a graph~$H=(V,E)$ and a vertex~$v\in V$, we denote the neighborhood of~$v$ by~$N_H(v) \coloneqq \{u\in V \mid \{u,v\} \in E\}$ and its degree by~$d_H(v) \coloneqq |N_H(v)|$.
Moreover, for a subset of vertices~$U\subseteq V$, let~$H[U]$ denote the subgraph induced by~$U$.
Our coloring argument uses \emph{Kempe exchanges}~\cite{Kempe1879}. 

\begin{definition}
    Let~$c\colon V\to[q]$ be a coloring of a graph~$H=(V,E)$, let~$a,b\in[q]$ be distinct colors, and let $H_{a,b}(c)\coloneqq H[\{v\in V\mid c(v)\in \{a,b\}\}]$.
    For a union~$W$ of connected components of~$H_{a,b}(c)$, the \emph{$(a,b)$-Kempe exchange on~$W$} interchanges the colors~$a$ and~$b$ on~$W$ and keeps all other colors.
\end{definition}
A Kempe exchange produces another valid coloring, since every edge leaving~$W$ has its other endpoint colored neither~$a$ nor~$b$.

For~$U\subseteq V$, we define the \emph{external boundary} of~$U$ in graph~$H=(V,E)$ by
\begin{align*}
    \partial_H U \coloneqq \{v\in V\setminus U\mid \{u,v\} \in E \text{ for some } u\in U \}.
\end{align*}
If the boundary avoids two colors, these colors can be interchanged inside~$U$.

\begin{lemma}
    \label{lem:boundary_kempe}
    Let~$c\colon V\to [q]$ be a coloring of~$H=(V,E)$, let~$U\subseteq V$, and let~$a,b\in~[q]$ be distinct colors that do not appear on~$\partial_H U$.
    Then interchanging~$a$ and~$b$ on~$U$ yields a coloring that is an $(a,b)$-Kempe exchange.
\end{lemma}
\begin{proof}
    Let~$W \coloneqq \{u\in U:c(u)\in\{a,b\}\}$.
    An edge of~$H_{a,b}(c)$ leaving~$W$ would end at a vertex of~$\partial_H U$ with color~$a$ or~$b$, which does not exist.
    Hence, $W$ is a union of components of~$H_{a,b}(c)$, so interchanging~$a$ and~$b$ on~$W$ is a Kempe exchange and produces a coloring.
\end{proof}

\section{Coloring the Conflict Graph} 
\label{sec:coloring-supercanonical-conflicts}

In this section, we prove the machine bound of~\cref{thm:main}.
Given a supercanonical schedule on~$m$ machines, we color its conflict graph with~$2m-1$ colors in strongly polynomial time.
In \Cref{sec:poly-time}, we show how to compute such a schedule.

Throughout this section, let~$\cS$ be a supercanonical schedule of~$J$ on~$m$ machines with canonical ordering~$\prec$.
Denote $G\coloneqq G^\cS$ and $N(k)\coloneqq N_G(k)$.

\subsection{Latest Common Blocks}

For adjacent jobs~$j$ and~$k$, define the last block in which both are processed by \begin{align*}
    L_{jk} \coloneqq \max\{t\in[T] \mid j,k \in \cS_t\}.
\end{align*}
We show that once two neighbors $k\prec j$ stop running together, job~$j$ does not run again until $k$ has completed.

\begin{lemma}
    \label{lem:latest_block}
    Let $k\in J$ and $j\in N(k)$ with $k\prec j$.
    Then $j\notin\cS_t$ for every block~$t$ with $L_{jk}<t\le\delta_k$.
\end{lemma}
\begin{proof}
    Suppose for contradiction that $j\in\cS_t$ for some block~$t$ with $L_{jk} < t \leq \delta_k$.
    Since $t$ and~$\delta_k$ both exceed~$L_{jk}$, maximality of~$L_{jk}$ gives $k\notin\cS_t$ and, as $k\in\cS_{\delta_k}$, also $j\notin\cS_{\delta_k}$.
    Hence $t\neq\delta_k$, so~$t<\delta_k$.
    Moreover, we have~$\rho_{k} \leq L_{jk} < t$.
    Thus, there is a priority swap for $k \prec j$ at blocks $t < \delta_k$, contradicting canonicity.
\end{proof}

\subsection{Separators at Pauses}
\label{subsec:separator}

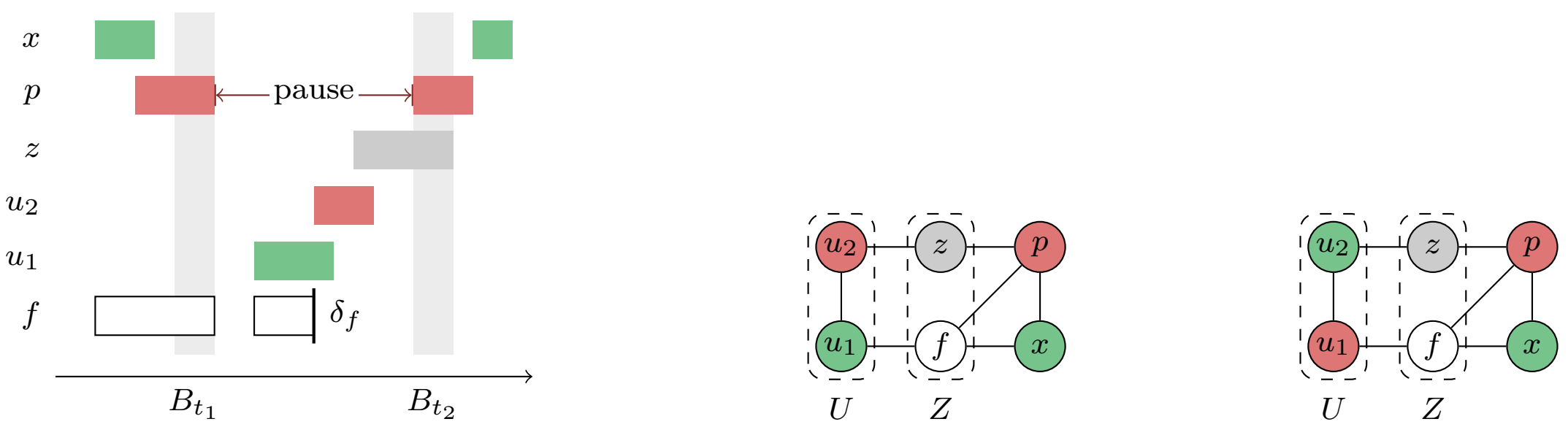
\begin{figure}[tb]
    \centering
    \begin{subfigure}[b]{0.46\textwidth}
        \centering
        \begin{tikzpicture}[x=0.36cm,y=0.5cm,every node/.style={font=\scriptsize}]
            \fill[gray!15] (3,-0.2) rectangle (4,6);
            \fill[gray!15] (9,-0.2) rectangle (10,6);
            \draw[fill=white] (1,0.15) rectangle (4,0.85);
            \draw[fill=white] (5,0.15) rectangle (6.5,0.85);
            \draw[thick] (6.5,0) -- (6.5,1) node[midway,right] {$\delta_f$};
            \fill[mygreen!70] (5,1.15) rectangle (7,1.85);
            \fill[myred!70] (6.5,2.15) rectangle (8,2.85);
            \fill[gray!40] (7.5,3.15) rectangle (10,3.85);
            \fill[myred!70] (2,4.15) rectangle (4,4.85);
            \fill[myred!70] (9,4.15) rectangle (10.5,4.85);
            \draw[myred!60!black,|<->|] (4,4.5) -- (9,4.5) node[midway,fill=white,inner sep=1pt,text=black] {pause};
            \fill[mygreen!70] (1,5.15) rectangle (2.5,5.85);
            \fill[mygreen!70] (10.5,5.15) rectangle (11.5,5.85);
            \draw[->] (0,-0.6) -- (12,-0.6);
            \node at (3.5,-1.1) {$B_{t_1}$};
            \node at (9.5,-1.1) {$B_{t_2}$};
            \foreach \y/\lab in {0/f,1/u_1,2/u_2,3/z,4/p,5/x}
                \node[left] at (0,\y+0.5) {$\lab$};
        \end{tikzpicture}
    \end{subfigure}
    \hfill
    \begin{subfigure}[b]{0.26\textwidth}
        \centering
        \begin{tikzpicture}[x=0.75cm,y=0.75cm,every node/.style={font=\scriptsize},
            vtx/.style={circle,draw,inner sep=0pt,minimum size=13pt}]
            \draw[dashed,rounded corners] (-0.4,-0.4) rectangle (0.4,1.6);
            \draw[dashed,rounded corners] (0.8,-0.4) rectangle (1.6,1.6);
            \node at (0,-0.75) {$U$};
            \node at (1.2,-0.75) {$Z$};
            \node[vtx,fill=mygreen!70] (u1) at (0,0)     {$u_1$};
            \node[vtx,fill=myred!70]   (u2) at (0,1.2)   {$u_2$};
            \node[vtx,fill=white]      (f)  at (1.2,0)   {$f$};
            \node[vtx,fill=gray!40]    (z)  at (1.2,1.2) {$z$};
            \node[vtx,fill=mygreen!70] (x)  at (2.4,0)   {$x$};
            \node[vtx,fill=myred!70]   (p)  at (2.4,1.2) {$p$};
            \draw (u1)--(u2) (u2)--(z) (z)--(p) (p)--(x)
                  (u1)--(f) (f)--(x) (f)--(p);
        \end{tikzpicture}
    \end{subfigure}
    \hfill
    \begin{subfigure}[b]{0.26\textwidth}
        \centering
        \begin{tikzpicture}[x=0.75cm,y=0.75cm,every node/.style={font=\scriptsize},
            vtx/.style={circle,draw,inner sep=0pt,minimum size=13pt}]
            \draw[dashed,rounded corners] (-0.4,-0.4) rectangle (0.4,1.6);
            \draw[dashed,rounded corners] (0.8,-0.4) rectangle (1.6,1.6);
            \node at (0,-0.75) {$U$};
            \node at (1.2,-0.75) {$Z$};
            \node[vtx,fill=myred!70]   (u1) at (0,0)     {$u_1$};
            \node[vtx,fill=mygreen!70] (u2) at (0,1.2)   {$u_2$};
            \node[vtx,fill=white]      (f)  at (1.2,0)   {$f$};
            \node[vtx,fill=gray!40]    (z)  at (1.2,1.2) {$z$};
            \node[vtx,fill=mygreen!70] (x)  at (2.4,0)   {$x$};
            \node[vtx,fill=myred!70]   (p)  at (2.4,1.2) {$p$};
            \draw (u1)--(u2) (u2)--(z) (z)--(p) (p)--(x)
                  (u1)--(f) (f)--(x) (f)--(p);
        \end{tikzpicture}
    \end{subfigure}
    \vspace*{-3ex}
    \caption{Schematic illustration of~\cref{lem:separator,lem:recoloring}, with colors~$a$ (red), $b$ (green), and other colors (gray), where~$f$ is uncolored.
    Left:~Job~$p$ pauses between blocks~$t_1$ and~$t_2$.
    The jobs in $U=\{u_1,u_2\}$ are processed only during the pause, and $Z=\{f,z\}$ contains the other jobs in~$B_{t_1}$ and~$B_{t_2}$.
    Color~$a$ last meets~$f$ in~$B_{t_1}$ via~$p$, while color~$b$ meets~$f$ later via~$u_1$.
    Middle:~The conflict graph with coloring~$c$, in which~$Z$ separates~$U$ from~$p$ and~$x$, and $Z'=\{z\}$ misses the colors~$a$ and~$b$.
    Right:~Exchanging~$a$ and~$b$ on~$U$ yields~$c'$, in which only~$x$ has color~$b$ among the neighbors of~$f$, 
    so $\lambda_b(c')<t_1=\lambda_a(c)$.}
    \label{fig:separator_recoloring}
\end{figure}

We show that whenever a job pauses, the jobs running together with it immediately before and after the pause separate the jobs processed entirely during the pause from the rest of the conflict graph.
A \emph{pause} of job~$k$ is a maximal nonempty sequence of blocks in which~$k$ does not run, preceded and followed by blocks in which it runs; see~\Cref{fig:separator_recoloring}.
Let~$t_1$ and~$t_2$ denote the blocks immediately before and after such a pause.
We define
\begin{align*}
    Z \coloneqq (\cS_{t_1} \cup \cS_{t_2}) \setminus\{k\}
    \qquad\text{and}\qquad
    U \coloneqq \{j\in J\mid [\rho_j,\delta_j] \subseteq (t_1,t_2)\}.
\end{align*}
Thus, $Z$ consists of the jobs sharing block~$t_1$ or~$t_2$ with job~$k$, and~$U$ consists of the jobs processed entirely during the pause.
In particular, $Z\cap U=\emptyset$ and $|Z|\le 2m-2$, since~$\cS_{t_1}$ and~$\cS_{t_2}$ each contain job~$k$ and at most $m-1$ other jobs.

\begin{lemma}
    \label{lem:separator}
    Every job processed in some block~$t$ with $t_1<t<t_2$ belongs to~$Z\cup U$.
    Consequently, the external boundary satisfies $\partial_G U\subseteq Z$.
\end{lemma}
\begin{proof}
    Let $j\in\cS_t$ for some $t_1<t<t_2$, and suppose that $j\notin Z$.
    Then $j\ne k$ and $j\notin\cS_{t_1}\cup\cS_{t_2}$.
    First, we show that $j\prec k$.
    Otherwise, $k\prec j$, and a priority swap exists for $k\prec j$ at blocks $t<t_2$, since $\rho_k\leq t_1 < t$, $j\in\cS_t\setminus\cS_{t_2}$, and $k\in\cS_{t_2}\setminus\cS_t$.
    Next, we show that $t_1<\rho_j$.
    Otherwise,~$\rho_j<t_1$ since~$j\notin\cS_{t_1}$, and a priority swap exists for $j\prec k$ at blocks $t_1 < t$, since $\rho_j < t_1$, $k\in\cS_{t_1}\setminus\cS_t$, and $j\in\cS_t\setminus\cS_{t_1}$.
    Finally, we show that $\delta_j<t_2$.
    Otherwise, $t_2<\delta_j$ since $j\notin \cS_{t_2}$.
    Then $\rho_k \leq t_1 < \rho_j \leq t < t_2 < \delta_j$, $k\notin\cS_{\rho_j}$, $k\in\cS_{t_2}$, and $j\notin\cS_{t_2}$, so a delay swap exists for~$k$, $j$, and block~$t_2$.
    Hence, $j\in U$.

    For~$\partial_G U\subseteq Z$, let $u\in U$ and $v\notin U$ be adjacent, and let~$t$ be a block with $u,v\in\cS_t$.
    Then $t_1<\rho_u\le t\le\delta_u<t_2$, so $v\in Z$ by the first statement.
\end{proof}

Thus, $Z$ separates~$U$ from $J\setminus(Z\cup U)$ in~$G$ (cf. \Cref{fig:separator_recoloring}).
By~\cref{lem:boundary_kempe}, any two colors absent from~$Z$ can therefore be interchanged on~$U$.

\subsection{Recoloring}

Let $q\coloneqq 2m-1$, let~$f$ be the first job in~$\prec$, and write $G-f \coloneqq G[J\setminus\{f\}]$.
Since deleting~$f$ preserves supercanonicity, we may assume by induction that~$G-f$ has a $q$-coloring.
We recolor it by Kempe exchanges until some color is missing on the neighborhood~$N(f)$, which we then assign to job~$f$.

Let~$c$ be a $q$-coloring of $G-f$ such that all~$q$ colors appear on~$N(f)$.
For each color~$a\in[q]$, define the last block in which an $a$-color job runs with~$f$ by
\begin{align*}
    \lambda_a(c) \coloneqq \max\{t\in[T] \mid f\in\cS_t \text{ and } c(j)=a \text{ for some } j\in\cS_t\setminus\{f\}\}.
\end{align*}
Moreover, let~$\Phi(c)\coloneqq\min_{a\in[q]}\lambda_a(c)$ be the earliest such block over all colors.
We choose a color whose last encounter with job~$f$ occurs earliest.
A local exchange then ensures that another color appears alongside~$f$ only before that block.

\begin{lemma}[Recoloring lemma]
    \label{lem:recoloring}
    Let~$c$ be a $q$-coloring of~$G-f$ in which all~$q$ colors appear on~$N(f)$.
    Then a Kempe exchange yields a $q$-coloring~$c'$ of~$G-f$ such that $|c'(N(f))|\le q-1$ or $\Phi(c')<\Phi(c)$.
\end{lemma}
\begin{proof}
    Let~$a$ be a color with $\lambda_a(c) = \Phi(c)$, let $t_1\coloneqq \lambda_a(c)$, and let $p\in\cS_{t_1}\setminus\{f\}$ with $c(p) = a$.
    Then~$L_{fp} = t_1$.
    Block~$\delta_f$ contains at most $m-1<q$ jobs besides~$f$, so some color does not appear there, and thus $t_1 < \delta_f$ by definition of~$\Phi$.
    Since $f\prec p$, we have $\delta_f \leq \delta_p$, and \Cref{lem:latest_block} gives $p\notin\cS_t$ for all $t_1<t\leq \delta_f$.
    Hence, job~$p$ has a next processing block~$t_2$ after~$t_1$, and $\delta_f<t_2$.

    Let~$Z$ and~$U$ be the sets of~\cref{subsec:separator} for the pause of~$p$ between~$t_1$ and~$t_2$, and let $Z'\coloneqq Z\setminus\{f\}$.
    Since $f\in\cS_{t_1}\setminus\cS_{t_2}$,
    \begin{align*}
        |Z'| = |(\cS_{t_1}\cup\cS_{t_2})\setminus\{f,p\}| \le (m-2)+(m-1) = q-2.
    \end{align*}
    All jobs in~$Z'$ are adjacent to~$p$, so color~$a$ is absent from~$Z'$.
    Furthermore, some color $b\ne a$ is also absent from~$Z'$.
    By~\cref{lem:separator}, $\partial_{G-f}U \subseteq Z'$, so interchanging~$a$ and~$b$ on~$U$ is a Kempe exchange by~\cref{lem:boundary_kempe}.
    Let~$c'$ be the resulting coloring, which agrees with~$c$ outside~$U$.

    Suppose that all~$q$ colors still appear on~$N(f)$ under~$c'$.
    Let~$j$ be a neighbor of~$f$ with $c'(j)=b$, and let~$t$ be a block with $f,j\in \cS_t$.
    We show that~$t<t_1$.
    If $t=t_1$, then $j\in \cS_{t_1}\setminus\{f,p\}\subseteq Z'$, since~$c'(p)=a$.
    This contradicts the absence of color~$b$ from~$Z'$, where~$c'$ agrees with~$c$.
    If $t > t_1$, then $t_1 < t \leq \delta_f < t_2$, so~\cref{lem:separator} gives $j\in Z'\cup U$.
    Again $j\notin Z'$, so $j\in U$, and hence $c(j) = a$.
    This contradicts $\lambda_a(c) = t_1$.
    Therefore every job of color~$b$ meets~$f$ only before~$t_1$, and consequently $\Phi(c') \leq \lambda_b(c') < t_1 = \Phi(c)$.
\end{proof}

We now prove that the power of migration is two.
Inductively applying \cref{lem:recoloring} yields the coloring in strongly polynomial time.

\begin{theorem}
    \label{thm:coloring}
    Given a supercanonical schedule~$\cS$ of~$n$ jobs on~$m$ machines with~$T$ blocks, a $(2m-1)$-coloring of the conflict graph~$G^\cS$ can be computed in time polynomial in~$n$ and~$T$, independently of the numerical input data.
\end{theorem}
\begin{proof}
    We may assume $m<n$ as otherwise each job could receive a distinct color.
    We use induction on~$n$.
    The base case~$n=0$ is trivial.
    Let~$f$ be the first job in~$\prec$.
    Since deleting~$f$ preserves supercanonicity, induction yields a $q$-coloring of~$G-f$.
    We apply~\cref{lem:recoloring} while all~$q$ colors appear on~$N(f)$, which frees a color or decreases the integer $\Phi\in[\rho_f,\delta_f]$.
    Hence, after at most~$T$ applications, some color is missing on~$N(f)$, which we assign to~$f$.

    After precomputing all~$\rho_j$ and~$\delta_j$ in $O(mT)$ time, each application of~\cref{lem:recoloring} takes $O(mT+n)$ time.
    Indeed, one scan over the blocks yields all~$\lambda_a$, the job~$p$, and the block~$t_2$, and the sets~$U$ and~$Z'$ and the exchange take $O(m+n)$ time.
    Since each of the~$n$ insertions requires at most~$T$ applications, the total running time is $O(nT(mT+n))$.
\end{proof}

\section{Constructing Supercanonical Schedules}
\label{sec:poly-time}

We show that every instance feasible on~$m$ machines admits a supercanonical schedule with~$O(n^2)$ blocks, computable in strongly polynomial time.
We first shrink windows to exclude delay swaps (\Cref{subsec:window-minimal}), then compute and realize a min-cost allocation to exclude priority swaps (\Cref{subsec:canonical_algorithm}).

\subsection{Window-Minimal Instances}
\label{subsec:window-minimal}

A delay swap postpones a job's first processing and thereby permits its window to shrink; see~\Cref{fig:swaps}. 
We therefore shrink windows as far as feasibility allows.
\begin{definition}
    \label{def:window_minimal}
    An instance is \emph{window-minimal} on~$m$ machines if it is feasible on~$m$ machines, but becomes infeasible when the window of any single job~$j$ is replaced by~$[R_j,D_j) \subsetneq [r_j,d_j)$.
\end{definition}
Any schedule of a window-minimal instance satisfies~$r_j = b_{\rho_j}$ and~$d_j = b_{\delta_j+1}$.

\begin{lemma}
    \label{lem:window_minimal}
    A schedule of a window-minimal instance admits no delay swap.
\end{lemma}
\begin{proof}
    Suppose that a delay swap exists for jobs~$j_1$ and $j_2$ at blocks~$\rho_{j_2}<t$.
    For $0 < \varepsilon\leq \min\{|B_{\rho_{j_2}}|,|B_t|\}$, we process~$j_1$ instead of~$j_2$ during $[b_{\rho_{j_2}}, b_{\rho_{j_2}}+\varepsilon)$ and~$j_2$ instead of~$j_1$ during~$[b_t, b_t+ \varepsilon)$.
    The number of running jobs and all processing amounts are unchanged, and no job runs twice at a time, since $j_1\notin \cS_{\rho_{j_2}}$ and $j_2\notin \cS_t$.
    Both new pieces lie in their windows, since $r_{j_1} \leq b_{\rho_{j_1}} < b_{\rho_{j_2}}$, $b_{t+1}\leq d_{j_1}$ by $j_1\in \cS_t$, and $b_{t+1} \leq b_{\delta_{j_2}}<d_{j_2}$.
    In the resulting feasible schedule, $j_2$ is processed only within $[r_{j_2} + \varepsilon,d_{j_2})$, contradicting window-minimality.
\end{proof}

Next, we compute window-minimal windows $[R_j,D_j)\subseteq[r_j,d_j)$.
Initially, $[R_j,D_j)\coloneqq[r_j,d_j)$.
We process the jobs once in a fixed order and, for each job~$j$, first maximize~$R_j$ and then minimize~$D_j$, keeping all other endpoints fixed.
Since later updates only shrink other windows, an endpoint that cannot be improved now cannot be improved later.

We maintain a set~$\mathcal A$ of event points containing all original release dates and deadlines and all endpoints introduced so far.
Let $a_1<\dots<a_\ell$ be its elements, and let $I_k\coloneqq[a_k,a_{k+1})$ and $\Delta_k\coloneqq a_{k+1}-a_k$ for $k\in[\ell-1]$.
We use the standard allocation formulation for preemptive feasibility~\cite{Horn74}, where $x_{jk}$ denotes the amount of processing of job~$j$ assigned to interval~$I_k$:
\begin{equation}\label{eq:allocation_lp}
    \begin{aligned}
        \textstyle\sum_{k\in [\ell-1]} x_{jk} &= p_j &&\forall j\in J,\\
        \textstyle\sum_{j\in J} x_{jk} &\leq m \Delta_k &&\forall k\in[\ell-1], \\
        0\le x_{jk} &\leq \Delta_k &&\forall j\in J,\ k\in[\ell-1], \\
        x_{jk} &= 0 &&\text{if } I_k\nsubseteq[R_j,D_j).
    \end{aligned}
    \tag{LP}
\end{equation}
This is a flow problem on the source-job-interval-sink network: source-job arcs have capacity~$p_j$, job-interval arcs have capacity~$\Delta_k$ and exist only if $I_k\subseteq[R_j,D_j)$, and interval-sink arcs have capacity~$m\Delta_k$.
A flow of value~$\sum_j p_j$ corresponds to a feasible allocation.
The windows $[R_j,D_j)$ are feasible on~$m$ machines if and only if~\eqref{eq:allocation_lp} is feasible.
Indeed, a schedule yields such an allocation, and McNaughton's wrap-around rule~\cite{McNaughton1959} realizes every feasible allocation interval by interval.
Placing a designated job~$j$ first in McNaughton's construction makes it run during $[a_k,a_k+x_{jk})$.
Reversing time within the interval instead places it during $[a_{k+1}-x_{jk},a_{k+1})$.

To maximize~$R_j$, we call a candidate $u\in\mathcal A\cap[R_j,D_j)$ \emph{feasible} if~\eqref{eq:allocation_lp} is feasible when~$j$ has window~$[u,D_j)$ and all other releases and deadlines are unchanged.
Feasibility is monotone in~$u$, so binary search finds the largest feasible candidate~$a_h$.
Let~$\beta$ be the minimum of~$x_{jh}$ subject to~\eqref{eq:allocation_lp} with window~$[a_h,D_j)$ for~$j$.
This is a min-cost flow problem with unit cost on arc~$(j,h)$ and zero cost on all other arcs.
We set $R_j\coloneqq a_{h+1}-\beta$ and insert it into~$\mathcal A$.
Symmetrically, to minimize~$D_j$, we find the smallest feasible candidate deadline~$a_{h+1}$, minimize~$x_{jh}$ subject to~\eqref{eq:allocation_lp} for window~$[R_j,a_{h+1})$, and set $D_j\coloneqq a_h+\beta$.

\begin{lemma}
    \label{lem:endpoint}
    Given all other windows, $a_{h+1}-\beta$ is the largest feasible release date of~$j$, and symmetrically, $a_h+\beta$ is the smallest feasible deadline of~$j$.
\end{lemma}
\begin{proof}
    We consider the release case, the deadline case is symmetric.
    By~\eqref{eq:allocation_lp}, we have $\beta\le\Delta_h$.
    We first show that $\beta>0$.
    If $a_{h+1}=D_j$, then~$I_h$ is the only interval in the window of~$j$, so $\beta=p_j>0$.   
    Otherwise, $\beta=0$ would make~$a_{h+1}$ a feasible candidate, contradicting the choice of~$a_h$.
    Placing~$j$ last in~$I_h$ realizes an optimal allocation with release date $a_{h+1}-\beta$, proving feasibility.
    Conversely, a feasible schedule with release date $u\in[a_h,a_{h+1})$ yields an allocation for window~$[a_h,D_j)$ with $\beta\le x_{jh}\le a_{h+1}-u$, so $u\le a_{h+1}-\beta$.
    Release dates $u\ge a_{h+1}$ are infeasible, by the choice of~$a_h$ and monotonicity if $a_{h+1}<D_j$, and since $p_j>0$ otherwise.
\end{proof}

\begin{theorem}
    \label{thm:window_minimal}
    Every instance feasible on $m$ machines admits a window-minimal instance satisfying $[R_j,D_j)\subseteq[r_j,d_j)$ for all $j\in J$.
    Such an instance can be computed in strongly polynomial time.
\end{theorem}
\begin{proof}
    Each minimization defining $\beta$ has a nonempty compact feasible region, so an optimum exists, also for real input data.
    By~\cref{lem:endpoint}, every update shrinks a window and preserves feasibility.
    Suppose that some final window $[R_j,D_j)$ could be replaced by a strict subinterval.
    Then we can increase~$R_j$ or decrease~$D_j$ while keeping all other final windows.
    When this endpoint was optimized, all windows contained the final ones.
    Hence, the resulting schedule would contradict the optimality of that endpoint at that step.

    Suppose that the input data is rational; scale it to integers.
    There are at most~$4n$ event points: the original endpoints and at most two new endpoints per job.
    The network therefore has $O(n)$ nodes and $O(n^2)$ arcs.
    Both feasibility testing and minimizing~$x_{jh}$ can be performed in strongly polynomial time~\cite{Orlin93}.
    With integral capacities, we can choose integral optimal flows.
    Thus, all endpoints remain integral and within the original horizon.
    Each of the~$2n$ updates solves $O(\log n)$ such~problems.
\end{proof}

\subsection{Constructing a Canonical Schedule}
\label{subsec:canonical_algorithm}
Let~$[R_j,D_j)$ be the window-minimal windows computed by~\cref{thm:window_minimal}, and let~$\mathcal A$, $I_k$, and~$\Delta_k$ be as in its last iteration.
By window-minimality, every feasible schedule completes job~$j$ at~$D_j$. Thus, the canonical ordering~$\prec$ is fixed by nondecreasing deadlines, with ties broken by job index.

Let $w_j\coloneqq n+1-\operatorname{rank}_\prec(j)$, where $\operatorname{rank}_\prec(j)\in[n]$ is the position of~$j$ in the canonical ordering~$\prec$.
We compute an optimal allocation~$x$ of
\begin{equation}\label{eq:canonical_lp}
    \min\ \Psi(x)\coloneqq \sum_{j\in J} \sum_{k\in [\ell-1]} w_j \cdot k \cdot x_{jk}
    \quad\text{subject to~\eqref{eq:allocation_lp}}.
\end{equation}
This is a min-cost flow problem on the network from~\Cref{subsec:window-minimal}, with cost~$w_j k$ on each job-interval arc~$(j,k)$.
The objective excludes priority swaps between different intervals, but does not determine the processing order within an interval.
We therefore realize each interval's allocation by a greedy rule that excludes priority swaps within that interval.

Fix an interval~$I_k$.
At time~$u\in I_k$, let $h\coloneqq a_{k+1}-u$ be the remaining length of~$I_k$, and let~$y_j$ be the remaining amount of~$x_{jk}$.
Initially, $u=a_k$ and $y_j=x_{jk}$.
A job is \emph{forced} if $y_j=h>0$, that is, it must run throughout the remainder of~$I_k$.
Initialize~$A$ with all forced jobs.
Add the earliest remaining jobs in the order~$\prec$ with~$y_j>0$, until $|A|=m$ or no such job remains.
We process~$A$ for
\begin{align*}
    \tau \coloneqq \min\bigl(\{h\} \cup \{y_j\mid j\in A\}\cup \{h-y_j \mid j\notin A,\ y_j>0\}\bigr),
\end{align*}
then increase~$u$ by~$\tau$, decrease~$y_j$ by~$\tau$ for all~$j\in A$, and repeat until $u=a_{k+1}$.
Thus, processing continues until an allocation finishes, an unselected job becomes forced, or the interval ends.

\begin{lemma}
    \label{lem:greedy}
    In every interval~$I_k$, the greedy rule realizes~$x$ with $O(n)$ blocks, and no priority swap exists for two blocks in~$I_k$.
\end{lemma}
\begin{proof}
    We maintain $0\leq y_j\leq h$ for every job~$j\in J$ and $\sum_{j\in J} y_j\leq mh$, which hold initially by~\eqref{eq:allocation_lp}.
    If $h>0$, at most~$m$ jobs are forced, so all of them belong to~$A$, and $|A|\leq m$.
    Hence, every job~$j\notin A$ with $y_j>0$ satisfies $y_j < h$, and thus $\tau>0$.
    For $j\in A$, both~$y_j$ and~$h$ decrease by~$\tau$, so $0\leq y_j\leq h$ is preserved by $\tau\leq y_j$.
    For $j\notin A$, only~$h$ decreases, so $y_j\leq h$ is preserved by $\tau\leq h-y_j$.
    If~$|A|=m$, both sides of $\sum_{j\in J} y_j\leq mh$ decrease by~$m\tau$.
    Otherwise, $A$ contains all jobs with $y_j>0$, so $\sum_{j\in J} y_j\leq |A|h\leq mh$.
    At $h=0$, we obtain $y_j=0$ for every job~$j\in J$, so~$x$ is realized in~$I_k$.

    A forced job remains forced until~$a_{k+1}$, as~$y_j$ and~$h$ decrease equally.
    Every iteration ending before~$a_{k+1}$ completes the allocation of a job or makes a job forced.
    Each happens at most once per job, so there are at most $2n+1$ iterations, each yielding one block.

    Suppose that a priority swap exists for jobs~$j_1\prec j_2$ at blocks~$s<t$ in~$I_k$.
    Since $j_1\in\cS_t$, we have~$y_{j_1}>0$ during block~$s$.
    As $j_1\notin\cS_s$ and $j_1\prec j_2$, the rule selects~$j_2$ in block~$s$ only if~$j_2$ is forced.
    Then~$j_2$ runs until~$a_{k+1}$, which contradicts~$j_2\notin\cS_t$.
\end{proof}

\begin{theorem}
    \label{thm:canonical}
    Every instance feasible on $m$ machines admits a feasible supercanonical schedule on $m$ machines with $O(n^2)$ blocks.
    Such a schedule can be computed in strongly polynomial time. 
\end{theorem}
\begin{proof}
    The allocation polytope is nonempty and compact, so an optimal allocation exists.
    By~\cref{lem:greedy}, realizing~$x$ greedily in every interval yields a feasible schedule~$\cS$ for the windows~$[R_j,D_j)$.
    Since $[R_j,D_j)\subseteq [r_j,d_j)$, schedule~$\cS$ is also feasible for the original instance.

    Suppose that a priority swap exists for~$j_1\prec j_2$ at blocks~$s<t$.
    By~\cref{lem:greedy}, we have~$B_s\subseteq I_k$ and $B_t\subseteq I_l$ with $k<l$.
    For $0<\varepsilon\le\min\{|B_s|,|B_t|\}$, we move~$\varepsilon$ of~$j_1$ from~$I_l$ to~$I_k$ and~$\varepsilon$ of~$j_2$ from~$I_k$ to~$I_l$.
    The capacity constraints remain satisfied, and so do the individual bounds, since~$j_1\notin\cS_s$ and~$j_2\notin\cS_t$.
    Moreover, $I_k$ lies in the window of~$j_1$, since $R_{j_1}\le b_{\rho_{j_1}}\le b_s$ with $R_{j_1}\in\mathcal A$, and~$D_{j_1}\ge a_{l+1}$.
    Similarly, $I_l$ lies in the window of~$j_2$, since $R_{j_2}\le a_k$ and $D_{j_2}\ge D_{j_1}\ge a_{l+1}$.
    Now, the resulting allocation~$x'$ is feasible, and contradicts optimality of~$x$, because
    \begin{align*}
        \Psi(x')-\Psi(x)=\varepsilon(w_{j_1}-w_{j_2})(k-l)<0.
    \end{align*}
    Thus, $\cS$ is canonical, and by~\cref{lem:window_minimal}, it is supercanonical.
    The schedule remains supercanonical when viewed as a schedule of the original instance, since supercanonicity does not depend on the windows.

    By~\cref{lem:greedy}, the greedy realization uses at most $2n+1$ blocks per interval, so~$\cS$ has at most~$(\ell-1)(2n+1)=O(n^2)$ blocks, as $\ell\le 4n$.
    The windows are computed in strongly polynomial time by~\cref{thm:window_minimal}.
    The min-cost flow problem can be solved in strongly polynomial time~\cite{Orlin93}. 
    After scaling the input to integers, we can choose an integral optimal allocation.
    Finally, the greedy rule performs $O(n)$ iterations per interval, each with $O(n)$ arithmetic operations.
\end{proof}

Combining \cref{thm:canonical}, which yields a supercanonical schedule with $O(n^2)$ blocks, with \cref{thm:coloring} proves \cref{thm:main}.
Thus, the power of migration for preemptive scheduling with time windows is two.

\section{Consequences and Extensions}
\label{sec:applications}

Our migration-elimination theorem also has consequences for other scheduling problems.
We discuss throughput maximization (\Cref{sec:throughput}), followed by speed augmentation and job partitioning (\Cref{sec:speed-augmentation}).
Finally, we consider identical machines with restricted assignment and show that a constant factor increase in the number of machines no longer suffices (\Cref{sec:restricted-assignment}).

\subsection{Throughput Maximization}
\label{sec:throughput}

In \emph{weighted preemptive throughput maximization}, each job has a nonnegative weight, and the goal is to select a maximum-weight subset of jobs that can be feasibly scheduled on~$m$ identical machines.
Let $\OPT_{\mathrm{mig}}(m)$ and $\OPT_{\mathrm{nm}}(m)$ denote the optimal values with and without migration, respectively.
Eliminating migration from an optimal migratory schedule by~\cref{thm:main} and retaining the~$m$ heaviest of the $2m-1$ machine classes yields
\begin{equation}\label{eq:throughput}
    \OPT_{\mathrm{nm}}(m)\geq \frac{m}{2m-1}\,\OPT_{\mathrm{mig}}(m).
\end{equation}
Koren, Dar, and Amir~\cite{KorenDA00} showed that this is tight, and conjectured~\eqref{eq:throughput} for weights proportional to processing times.
Our result confirms this conjecture, even for arbitrary weights.
Combining our theorem with repeated applications of the single-machine FPTAS in~\cite{PruhsW07} gives the following approximation guarantee against the migratory optimum.

\begin{corollary}
    \label{cor:throughput-approx}
    For every $\delta>0$, weighted preemptive throughput maximization on $m$ identical machines admits a polynomial-time non-migratory algorithm with approximation ratio
    \[
        \left[1-\left(1-\frac{1}{2m-1}\right)^m\right]^{-1}+\delta
        \leq (1-e^{-1/2})^{-1}+\delta<2.542+\delta
    \]
    against the migratory optimum.
\end{corollary}
\begin{proof}
    We may assume $m\leq n$, since otherwise each individually feasible job gets its own machine.
    Apply the single-machine FPTAS~\cite{PruhsW07} repeatedly to the remaining jobs, scheduling the selected jobs on one new machine in each of $m$ iterations, as in the reduction of~\cite{KalP01}.
    Write $W_i$ for the weight selected after $i$ iterations, with $W_0=0$, and $O=\OPT_{\mathrm{mig}}(m)$.
    The unselected jobs of a migratory optimum have weight at least $O-W_i$ and, by \Cref{thm:main}, partition into $2m-1$ feasible single-machine classes.
    Thus, using the FPTAS with accuracy parameter $\eps\in(0,1)$ and approximation guarantee~$1-\eps$,
    \[
        W_{i+1}-W_i\geq\frac{1-\eps}{2m-1}(O-W_i).
    \]
    Iterating this recurrence yields
    \[
        W_m\geq\left[1-\left(1-\frac{1-\eps}{2m-1}\right)^m\right]O.
    \]
    Choosing $\eps>0$ sufficiently small as a function of $\delta$ gives the first bound.
    The uniform bound follows from $\left(1-\frac{1}{2m-1}\right)^m\leq e^{-m/(2m-1)}\leq e^{-1/2}$.
\end{proof}

For weights proportional to processing times, Koren, Dar, and Amir~\cite{KorenDA00} use the same iterative scheme to obtain a guarantee tending to $1-e^{-1/2}$.
\Cref{cor:throughput-approx} extends this guarantee to arbitrary weights and improves the bound for each fixed~$m$ and sufficiently small~$\delta$.
For arbitrary weights, it improves the approximation ratio $12+\delta$ obtained by combining the Kalyanasundaram--Pruhs reduction~\cite{KalP01} with the single-machine FPTAS, as reported in~\cite{KarakostasK25}.

For online throughput maximization, \eqref{eq:throughput} reduces the loss in transferring guarantees to the migratory optimum from $6-5/m$~\cite{KalP01} to $2-1/m$, improving the constants in~\cite{EberleMS23,Eberle24}.
More generally, retaining $k\leq m$ machine classes gives $\OPT_{\mathrm{nm}}(k)\geq\frac{k}{2m-1}\,\OPT_{\mathrm{mig}}(m)$, improving the fraction~$k/(6m)$ in the machine-reduction argument of~\cite{MoseleyPSZ24}.

\subsection{Speed Augmentation and Job Partitioning}
\label{sec:speed-augmentation}

Instead of additional machines, \cref{thm:main} also gives an asymptotically optimal speedup bound for eliminating migration on the original number of machines.

\begin{corollary}
\label{cor:speed-two}
    Every instance feasible with migration on~$m$ identical unit-speed machines can be scheduled without migration on~$m$ speed-two machines in strongly polynomial time. 
    The speedup factor is asymptotically optimal.
\end{corollary}
\begin{proof}
    By~\cref{thm:main}, we compute a feasible non-migratory schedule on $2m-1$ unit-speed machines in strongly polynomial time.
    We add an idle machine and group the $2m$ machines into~$m$ pairs, each simulated by one machine of speed~$2$.
    Consider a block~$B_t$ of length~$\ell$.
    On each machine of a pair, at most one job is processed throughout~$B_t$.
    The speed-$2$ machine processes the job of the first machine during the first half of~$B_t$ and the job of the second machine during the second half.
    Each job thus receives the same processing amount within~$B_t$, so all processing times and windows are preserved, and every job is processed on a single machine.
    The number of blocks at most doubles.

    For the lower bound, consider $m+1$ jobs with common window $[0,1)$ and processing time $m/(m+1)$~\cite[Section~5]{ChanLT05}.
    McNaughton's wrap-around rule~\cite{McNaughton1959} schedules them migratorily on~$m$ unit-speed machines.
    In every non-migratory schedule on~$m$ machines, some machine processes two jobs, which requires speed at least $2m/(m+1)$.
\end{proof}

Equivalently, let $s_{\mathrm{mig}}(J,m)$ and $s_{\mathrm{nm}}(J,m)$ denote the minimum common speed of~$m$ machines needed to schedule~$J$ with and without migration, respectively.
Since running at speed~$s$ is equivalent to dividing all processing times by~$s$, \cref{cor:speed-two} yields
\[
    s_{\mathrm{mig}}(J,m) \leq s_{\mathrm{nm}}(J,m) \leq 2\cdot s_{\mathrm{mig}}(J,m).
\]

An earlier speedup bound of~$4-1/m$ follows implicitly from the non-migratory feasibility condition of Baruah and Fisher~\cite[Corollary~1]{BaruahFisher_RTS2008_NonMigratory_Feasibility}.
Their condition was derived in the context of multiprocessor real-time scheduling of recurrent tasks, including periodic and sporadic task systems, and applies to arbitrary collections of jobs.
Indeed, migratory feasibility bounds the total processing requirement of jobs whose windows lie in any interval by~$m$ times the interval length, and each job's density~$p_j/(d_j-r_j)$ by~$1$.
Running at speed~$4-1/m$ therefore reduces the load and maximum job density to at most~$m^2/(4m-1)$ and~$m/(4m-1)$, respectively, which are precisely the sufficient thresholds in their corollary.
\Cref{cor:speed-two} improves this guarantee to~$2$.

Goossens, Richard, Lindström, Lupu, and Ridouard~\cite{GoossensRLLR12} study offline \emph{job partitioning}, which assigns each job to one processor before execution. 
Our results give machine and speed augmentation guarantees for this model. 
The stronger notion of \emph{task partitioning}, usually called partitioned scheduling in the real-time literature~\cite{EkbergB21}, requires all jobs of the same recurrent task to use the same processor.
Our construction may assign different jobs of a task to different processors and therefore gives no guarantee for task partitioning.

\subsection{A Lower Bound for Restricted Assignment}
\label{sec:restricted-assignment}

The preceding guarantees rely on every job being eligible on every machine.
We now show that even two eligible machines per job can prevent elimination of migration with a constant factor increase in the number of machines.

Under restricted assignment, each job~$j$ may only be processed on the machines in its eligible set $M_j\subseteq[m]$.
To eliminate migration, we use additional copies of the machines, where a copy of machine~$i$ may only process jobs~$j$ with $i\in M_j$.

There are two natural ways to count copies.
In the \emph{uniform} model, every machine has the same number~$\mu$ of copies, so $\mu m$ machines are used in total.
In the \emph{free} model, the number of copies may be chosen separately for every machine, and we count the total number of copies, including the original machines.
The free model is more permissive, so a lower bound of~$N$ copies in total in the free model implies $\mu\geq N/m$ in the uniform model.

In the uniform model, Im and Moseley~\cite{ImM16} show that speed $1+\varepsilon$ and $\mu=O(1/\varepsilon^2)$ suffice to eliminate migration on related machines with restricted assignment, whereas without additional speed, no constant~$\mu$ suffices on related machines.
Their lower bound relies on uniform copies: with free copies, copying only the fast machines yields $O(m)$ copies in total.
We show that under unit-speed restricted assignment with at most two eligible machines per job, $\Omega(m\log m)$ copies in total may be necessary even in the free model, and thus $\mu=\Omega(\log m)$ in the uniform model.

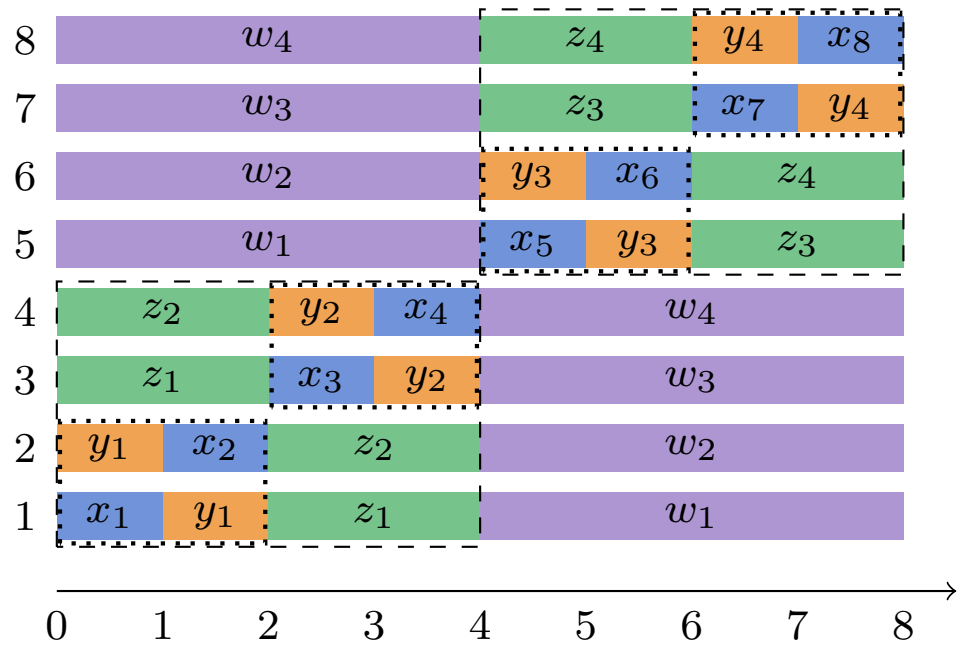
\begin{figure}[t]
    \centering
    \begin{tikzpicture}[x=0.7cm,y=0.45cm,every node/.style={font=\scriptsize}]
        \foreach \m/\s/\i in {1/0/1,2/1/2,3/2/3,4/3/4,5/4/5,6/5/6,7/6/7,8/7/8} {
            \fill[myblue!70] (\s,\m-0.85) rectangle ({\s+1},\m-0.15);
            \node at ({\s+0.5},\m-0.5) {$x_{\i}$};
        }
        \foreach \m/\s/\i in {2/0/1,1/1/1,4/2/2,3/3/2,6/4/3,5/5/3,8/6/4,7/7/4} {
            \fill[myorange!80] (\s,\m-0.85) rectangle ({\s+1},\m-0.15);
            \node at ({\s+0.5},\m-0.5) {$y_{\i}$};
        }
        \foreach \m/\s/\i in {3/0/1,1/2/1,4/0/2,2/2/2,7/4/3,5/6/3,8/4/4,6/6/4} {
            \fill[mygreen!70] (\s,\m-0.85) rectangle ({\s+2},\m-0.15);
            \node at ({\s+1},\m-0.5) {$z_{\i}$};
        }
        \foreach \m/\s/\i in {5/0/1,1/4/1,6/0/2,2/4/2,7/0/3,3/4/3,8/0/4,4/4/4} {
            \fill[mypurple!60] (\s,\m-0.85) rectangle ({\s+4},\m-0.15);
            \node at ({\s+2},\m-0.5) {$w_{\i}$};
        }
        \draw[dashed] (0,0.05) rectangle (4,3.95);
        \draw[dashed] (4,4.05) rectangle (8,7.95);
        \foreach \a/\b in {0/0,2/2,4/4,6/6}
            \draw[dotted,thick] (\a+0.03,\b+0.1) rectangle (\a+1.97,\b+1.9);
        \draw[->] (0,-0.6) -- (8.5,-0.6);
        \foreach \t in {0,...,8}
            \node at (\t,-1.1) {$\t$};
        \foreach \i in {1,...,8}
            \node[left] at (0,\i-0.5) {$\i$};
    \end{tikzpicture}
    \caption{A migratory schedule of the instance~$I_3$ of~\cref{prop:restricted_assignment} on eight machines, with machines as rows and time on the horizontal axis.
    The dashed and dotted boxes contain the copies of~$I_2$ and~$I_1$, respectively, the jobs~$x_i$ form the copies of~$I_0$, and the jobs~$y_q$, $z_q$, and~$w_i$ are added on levels~$1$, $2$, and~$3$.
    Each job is processed exactly on its eligible machines, and its processing time equals its window length.
    Without migration, each of the $8+3\cdot 4=20$ jobs requires its own machine copy.}
    \label{fig:restricted_assignment}
\end{figure}

\begin{proposition}
    \label{prop:restricted_assignment}
    Let $m\geq1$ and $k\coloneqq\lfloor\log_2 m\rfloor$.
    Under unit-speed restricted assignment, eliminating migration may require $m+k2^{k-1}$ machine copies in total, even if every job is eligible on at most two machines.
    In particular, $\Omega(m\log m)$ copies in total and $\mu=\Omega(\log m)$ in the uniform model may be necessary.
\end{proposition}
\begin{proof}
    We recursively construct an instance~$I_k$ on the machines~$[2^k]$ with all windows in $[0,2^k)$, and leave the remaining $m-2^k$ machines idle.
    Let~$I_0$ consist of a single job with processing time~$1$, window $[0,1)$, and eligible machine~$1$.
    For~$k\geq1$, we let $h\coloneqq 2^{k-1}$.
    We take one copy of~$I_{k-1}$ on the machines $1,\dots,h$, and a second copy on the machines $h+1,\dots,2h$ with all windows shifted by~$h$.
    Moreover, for every $i\in[h]$, we add a job~$z_i$ with processing time~$2h$, window~$[0,2h)$, and eligible machines $M_{z_i}=\{i,i+h\}$ (see~\cref{fig:restricted_assignment}).
    Thus, every job is eligible on at most two machines, and the number of jobs satisfies $|I_k|=2|I_{k-1}|+2^{k-1}$ with $|I_0|=1$, so $|I_k|=2^k+k2^{k-1}$.

    We show by induction on~$k$ that~$I_k$ is feasible with migration on~$2^k$ machines.
    We schedule both copies of~$I_{k-1}$ recursively, and we process each job~$z_i$ on machine~$i+h$ during $[0,h)$ and on machine~$i$ during $[h,2h)$.
    This is feasible, since the jobs of the first copy only use the machines $1,\dots,h$ during~$[0,h)$, and the jobs of the second copy only use the machines $h+1,\dots,2h$ during~$[h,2h)$.

    It remains to show that every non-migratory schedule of~$I_k$ uses a separate copy for each job.
    The recursion assigns to each machine~$i\in M$ and each level $\ell\in\{0,\dots,k\}$ exactly one copy of~$I_\ell$ that contains machine~$i$, and the windows of these copies are nested intervals.
    Every job eligible on machine~$i$ is added in one of these copies and has its window, so any two jobs with a common eligible machine have overlapping windows.
    Moreover, every job has processing time equal to its window length, so in a non-migratory schedule, it occupies its machine copy throughout its window.
    Hence, no machine copy processes two jobs.
    Together with the $m-2^k$ idle machines, every non-migratory schedule uses at least $|I_k|+m-2^k=m+k2^{k-1}$ copies in total.
    Since $2^{k-1} > m/4$ and $k > \log_2 m-1$, this is $\Omega(m\log m)$, and the uniform model requires $\mu\geq(m+k2^{k-1})/m = \Omega(\log m)$.
\end{proof}

Whether $O(\log m)$ copies of each machine always suffice under unit-speed restricted assignment remains open.

\section*{AI disclosure} We used ChatGPT-6 Astra and Claude Opus 5.5 for mathematical assistance, language editing, and figure preparation.

We established the~tight $3m-2$ coloring bound for canonical schedules~(\Cref{appendix:canonical-3m-2}), introduced supercanonical schedules, and established the limitations of the minimum-degree approach~(\Cref{appendix:supercanonical_min_degree}).
We developed the swap framework, proved the latest-block lemma~(\Cref{lem:latest_block}) and an earlier, weaker separator lemma, and proposed the inductive coloring strategy using Kempe exchanges.
We also proposed the rank-weighted LP objective for constructing canonical schedules and simplified the coloring potential to~$\Phi(c) = \min_a \lambda_a(c)$.

Building on this framework, ChatGPT-6 Astra developed the stronger separator lemma~(\Cref{lem:separator}), the quantities $\lambda_a(c)$, an initial sum-based potential, and the specific recoloring construction that completed the coloring argument~(\Cref{lem:recoloring}).
It also provided the polynomial-time construction of a window-minimal instance.

We verified all AI-assisted contributions and take full responsibility for the content and correctness of the paper. 

\bibliographystyle{plain}
\bibliography{literature.bib}

\appendix

\section*{Appendices}

\section{Canonical Schedules: A Tight Coloring Bound}
\label[appendix]{appendix:canonical-3m-2}

In this appendix, we revisit the approach of Kalyanasundaram and Pruhs~\cite{KalP01}, who use a density argument to bound the minimum degree of the conflict graph of a canonical schedule.
For a nonempty graph~$H=(V,E)$, the minimum degree of~$H$ is~$\delta(H) \coloneqq \min_{v\in V} d_H(v)$.
They then apply the classical Szekeres--Wilf bound~\cite{SzekeresW68}: for every graph $G=(V,E)$,
\begin{align}
    \label{eq:szekeres_wilf}
    \chi(G) &\leq 1 + \max_{\emptyset \neq U\subseteq V} \delta(G[U]).
\end{align}
The bound follows by repeatedly removing a vertex of minimum degree and then greedily coloring the vertices in reverse order.
Each vertex has at most $\max_{\emptyset\neq U\subseteq V}\delta(G[U])$ already colored neighbors, so a palette of one more color suffices.
Thus, to bound the chromatic number of the conflict graph, it suffices to find a job of small degree in every nonempty induced subgraph.

By bounding the density of a supergraph of the conflict graph by~$3(m-1)$, \cite{KalP01} obtain a job of degree at most~$6m-6$, and thus a non-migratory schedule on~$6m-5$ machines.
By bounding the degree of a suitable job directly, we improve this to~$3m-2$ machines.
However, this bound is tight, as there are canonical schedules whose conflict graph is a clique on~$3m-2$ jobs.
Hence, canonicity alone does not suffice for~$2m-1$ machines, which motivates supercanonical schedules and the recoloring argument of~\cref{sec:coloring-supercanonical-conflicts}.

\begin{theorem}
    \label{thm:canonical_3m-2}
    Let~$\cS$ be a canonical schedule on~$m$ machines.
    Then the chromatic number of the conflict graph satisfies $\chi(G^\cS)\le 3m-2$, and this bound is tight.
\end{theorem}

We prove the upper bound in~\cref{app:structure-lemma,app:ngh-decomposition,app:root-sequence} and its tightness in \cref{app:canonical-tightness}.
In more detail, we identify in~\cref{app:structure-lemma} a block in which many neighbors of a job~$k$ must be processed.
Next, in~\cref{app:ngh-decomposition}, we partition the neighborhood of~$k$ accordingly and bound most of its parts.
Finally, in~\cref{app:root-sequence}, we find a job for which the remaining parts are small as well.

Throughout, let~$\cS$ be a canonical schedule on~$m$ machines with canonical ordering~$\prec$, let $G\coloneqq G^\cS$, and let $N(k)\coloneqq N_G(k)$.
We use the span $[\rho_j,\delta_j]$ of a job~$j$ in place of its window, that is,~$r_j \coloneqq b_{\rho_j}$ and~$d_j \coloneqq b_{\delta_j+1}$.
Moreover, \cref{lem:latest_block} only uses canonicity, so it also applies here.

\subsection{Structure Lemma}
\label{app:structure-lemma}

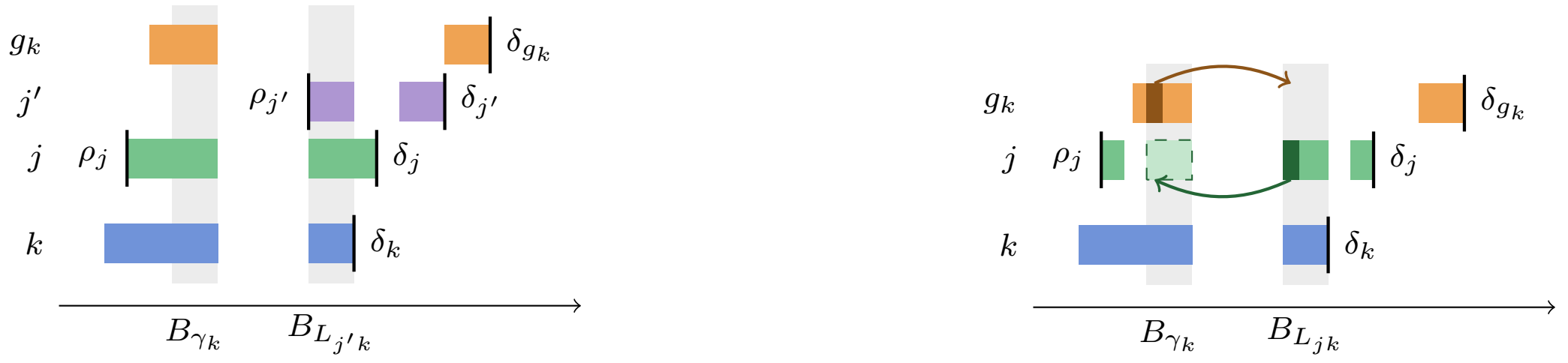
\begin{figure}[t]
    \centering
    \begin{subfigure}[b]{0.48\textwidth}
        \centering
        \begin{tikzpicture}[x=0.4cm,y=0.5cm,every node/.style={font=\scriptsize}]
            \fill[gray!15] (2.5,-0.2) rectangle (3.5,4.7);
            \fill[gray!15] (5.5,-0.2) rectangle (6.5,4.7);
            \fill[myblue!70] (1,0.15) rectangle (3.5,0.85);
            \fill[myblue!70] (5.5,0.15) rectangle (6.5,0.85);
            \draw[thick] (6.5,0) -- (6.5,1) node[midway,right] {$\delta_k$};
            \fill[mygreen!70] (1.5,1.65) rectangle (3.5,2.35);
            \fill[mygreen!70] (5.5,1.65) rectangle (7,2.35);
            \draw[thick] (1.5,1.5) -- (1.5,2.5) node[midway,left] {$\rho_j$};
            \draw[thick] (7,1.5) -- (7,2.5) node[midway,right] {$\delta_j$};
            \fill[mypurple!60] (5.5,2.65) rectangle (6.5,3.35);
            \fill[mypurple!60] (7.5,2.65) rectangle (8.5,3.35);
            \draw[thick] (5.5,2.5) -- (5.5,3.5) node[midway,left] {$\rho_{j'}$};
            \draw[thick] (8.5,2.5) -- (8.5,3.5) node[midway,right] {$\delta_{j'}$};
            \fill[myorange!80] (2,3.65) rectangle (3.5,4.35);
            \fill[myorange!80] (8.5,3.65) rectangle (9.5,4.35);
            \draw[thick] (9.5,3.5) -- (9.5,4.5) node[midway,right] {$\delta_{g_k}$};
            \draw[->] (0,-0.6) -- (11.5,-0.6);
            \node at (3,-1.1) {$B_{\gamma_k}$};
            \node at (6,-1.1) {$B_{L_{j'k}}$};
            \foreach \y/\lab in {0.5/k,2/j,3/j',4/g_k}
                \node[left] at (0,\y) {$\lab$};
        \end{tikzpicture}
    \end{subfigure}
    \hfill
    \begin{subfigure}[b]{0.48\textwidth}
        \centering
        \begin{tikzpicture}[x=0.4cm,y=0.5cm,every node/.style={font=\scriptsize}]
            \fill[gray!15] (2.5,-0.2) rectangle (3.5,3.7);
            \fill[gray!15] (5.5,-0.2) rectangle (6.5,3.7);
            \fill[myblue!70] (1,0.15) rectangle (3.5,0.85);
            \fill[myblue!70] (5.5,0.15) rectangle (6.5,0.85);
            \draw[thick] (6.5,0) -- (6.5,1) node[midway,right] {$\delta_k$};
            \fill[mygreen!70] (1.5,1.65) rectangle (2,2.35);
            \fill[mygreen!30] (2.5,1.65) rectangle (3.5,2.35);
            \draw[mygreen!60!black,dashed] (2.5,1.65) rectangle (3.5,2.35);
            \fill[mygreen!70] (5.5,1.65) rectangle (6.5,2.35);
            \fill[mygreen!60!black] (5.5,1.65) rectangle (5.85,2.35);
            \fill[mygreen!70] (7,1.65) rectangle (7.5,2.35);
            \draw[thick] (1.5,1.5) -- (1.5,2.5) node[midway,left] {$\rho_j$};
            \draw[thick] (7.5,1.5) -- (7.5,2.5) node[midway,right] {$\delta_j$};
            \fill[myorange!80] (2.2,2.65) rectangle (3.5,3.35);
            \fill[myorange!60!black] (2.5,2.65) rectangle (2.85,3.35);
            \fill[myorange!80] (8.5,2.65) rectangle (9.5,3.35);
            \draw[thick] (9.5,2.5) -- (9.5,3.5) node[midway,right] {$\delta_{g_k}$};
            \draw[->,thick,mygreen!60!black] (5.675,1.65) to[bend left=25] (2.675,1.65);
            \draw[->,thick,myorange!60!black] (2.675,3.35) to[bend left=25] (5.675,3.35);
            \draw[->] (0,-0.6) -- (11.5,-0.6);
            \node at (3,-1.1) {$B_{\gamma_k}$};
            \node at (6,-1.1) {$B_{L_{jk}}$};
            \foreach \y/\lab in {0.5/k,2/j,3/g_k}
                \node[left] at (0,\y) {$\lab$};
        \end{tikzpicture}
    \end{subfigure}
    \caption{Schematic illustration of~\cref{lem:structure}.
    Left:~The critical block~$B_{\gamma_k}$ is the earliest last common block of~$k$ with a job in~$N_0(k)$, attained by the pivot~$g_k$, while $L_{j'k}>\gamma_k$.
    The neighbor~$j\prec g_k$ satisfies $\rho_j\le\gamma_k\le\delta_j$ and is processed in~$B_{\gamma_k}$.
    Right:~If~$j$ were not processed in~$B_{\gamma_k}$ (dashed), then~$j$ runs in a later common block~$B_{L_{jk}}$ with~$k$, in which~$g_k$ does not run.
    Exchanging the dark pieces is then a priority swap for $j\prec g_k$ at blocks $\gamma_k<L_{jk}$.}
    \label{fig:structure}
\end{figure}

Fix a job~$k$ with $|N(k)|\ge m$, as jobs with fewer neighbors have degree at most $m-1\le 3m-3$.
Let~$N_0(k)$ be the set of the~$m$ last neighbors of~$k$ in~$\prec$.
We define the \emph{critical block} of~$k$ as the earliest last common block of~$k$ with a job in~$N_0(k)$, that is,
\begin{align*}
    \gamma_k\coloneqq\min\{L_{j'k}\mid j'\in N_0(k)\}.
\end{align*}
Moreover, we fix a \emph{pivot} $g_k\in N_0(k)$ with $L_{g_k k}=\gamma_k$.
Thus, both~$k$ and~$g_k$ are processed in block~$\gamma_k$, and $L_{j'k}\ge\gamma_k$ for every $j'\in N_0(k)$.
Every neighbor outside~$N_0(k)$ precedes the pivot.
The following lemma therefore places all such neighbors whose spans contain~$\gamma_k$ in the same processing block, where the machine capacity bounds their number.
The following structure lemma states that every neighbor of~$k$ that precedes the pivot and whose span contains the critical block is processed in the critical block (see~\Cref{fig:structure}).

\begin{lemma}
    \label{lem:structure}
    Let $j\in N(k)$ with $j\prec g_k$ and $\rho_j\le\gamma_k\le\delta_j$.
    Then $j\in\cS_{\gamma_k}$.
\end{lemma}
\begin{proof}
    It suffices to show that $L_{jk}\ge\gamma_k$.
    Indeed, if $L_{jk}=\gamma_k$, then~$j\in\cS_{\gamma_k}$.
    If $L_{jk}>\gamma_k$, then~$g_k\notin\cS_{L_{jk}}$ by maximality of~$L_{g_k k}=\gamma_k$, since $k\in\cS_{L_{jk}}$.
    Hence, if $j\notin\cS_{\gamma_k}$, a priority swap exists for $j\prec g_k$ at blocks $\gamma_k<L_{jk}$, since $\rho_j\le\gamma_k$, $g_k\in\cS_{\gamma_k}\setminus\cS_{L_{jk}}$, and $j\in\cS_{L_{jk}}\setminus\cS_{\gamma_k}$.
    We show $L_{jk}\ge\gamma_k$ by distinguishing three cases.

    Case 1: Suppose that $\delta_j\le\delta_k$, and assume for contradiction that $L_{jk}<\gamma_k$.
    If $j\prec k$, then~\cref{lem:latest_block} gives $k\notin\cS_{\gamma_k}$, since $L_{jk}<\gamma_k\le\delta_j$.
    If $k\prec j$, then~\cref{lem:latest_block} gives $j\notin\cS_{\delta_j}$, since $L_{jk}<\delta_j\le\delta_k$.
    Both contradict the definitions of~$\gamma_k$ and~$\delta_j$.
    
    Case 2: Suppose that $\delta_j>\delta_k$ and $j\in N_0(k)$.
    Then $L_{jk}\ge\gamma_k$ by definition of the critical block~$\gamma_k$.
    
    Case 3: Suppose that $\delta_j>\delta_k$ and $j\notin N_0(k)$.
    Then $j\prec j'$ for every $j'\in N_0(k)$, and $\rho_j\le\gamma_k\le L_{j'k}\le\delta_k<\delta_j$.
    If $j\notin\cS_{L_{j'k}}$, then $j'\in\cS_{\delta_j}$, as otherwise a priority swap exists for $j\prec j'$ at blocks $L_{j'k}<\delta_j$.
    Since~$\cS_{\delta_j}$ contains~$j$ and at most $m-1$ other jobs, we obtain $j\in\cS_{L_{j'k}}$ for some~$j'\in N_0(k)$.
    Thus, $L_{jk}\ge L_{j'k}\ge\gamma_k$.
\end{proof}

\subsection{Neighborhood Decomposition}
\label{app:ngh-decomposition}

We partition the neighbors of~$k$ outside~$N_0(k)$ according to the position of their spans relative to the critical block~$\gamma_k$ and the span of~$k$.
Note that $\rho_k\le\gamma_k\le\delta_k$, since~$k$ is processed in block~$\gamma_k$.
We define
\begin{align*}
    N_1(k) &\coloneqq \{j\in N(k)\setminus N_0(k)\mid \rho_j\le\gamma_k\le\delta_j\},\\
    N_2(k) &\coloneqq \{j\in N(k)\setminus N_0(k)\mid \rho_j<\rho_k \text{ and } \delta_j<\gamma_k\},\\
    N_3(k) &\coloneqq \{j\in N(k)\setminus N_0(k)\mid \gamma_k<\rho_j \text{ and } \delta_k<\delta_j\},\\
    N_4(k) &\coloneqq \{j\in N(k)\setminus N_0(k)\mid \gamma_k\notin[\rho_j,\delta_j] \text{ and } [\rho_j,\delta_j]\subseteq[\rho_k,\delta_k]\}.
\end{align*}
Thus, the spans of jobs in~$N_1(k)$ contain the critical block.
The spans of jobs in~$N_2(k)$ lie before the critical block and start before~$k$, and the spans of jobs in~$N_3(k)$ lie after the critical block and end after~$k$.
The spans of jobs in~$N_4(k)$ avoid the critical block and lie within the span of~$k$.

\begin{figure}[t]
    \centering
    \begin{tikzpicture}[x=0.45cm,y=0.5cm,every node/.style={font=\scriptsize}]
        \fill[gray!15] (4,-0.2) rectangle (5,7.2);
        \draw[dashed,gray] (2,-0.2) -- (2,7.2);
        \draw[dashed,gray] (9,-0.2) -- (9,7.2);
        \fill[mypurple!60] (0.5,0.3) rectangle (3,0.7);
        \fill[gray!50] (2.5,1.3) rectangle (3.75,1.7);
        \fill[mygreen!70] (1,2.3) rectangle (6,2.7);
        \fill[gray!50] (6,3.3) rectangle (8,3.7);
        \fill[myblue!70] (2,4.3) rectangle (9,4.7);
        \fill[myred!60] (6,5.3) rectangle (10.5,5.7);
        \fill[myorange!80] (3.5,6.3) rectangle (11.5,6.7);
        \draw[->] (0,-0.6) -- (12.5,-0.6);
        \node at (2,-1.1) {$\rho_k$};
        \node at (4.5,-1.1) {$B_{\gamma_k}$};
        \node at (9,-1.1) {$\delta_k$};
        \foreach \y/\lab in {0/N_2(k),1/N_4(k),2/N_1(k),3/N_4(k),4/k,5/N_3(k),6/N_0(k)}
            \node[left] at (0,\y+0.5) {$\lab$};
    \end{tikzpicture}
    \caption{Schematic illustration of~\cref{lem:decomposition}, showing spans instead of processing: the bar of a job~$j$ extends from the start of~$B_{\rho_j}$ to the end of~$B_{\delta_j}$.
    The spans in~$N_1(k)$ contain the critical block~$\gamma_k$.
    The spans in~$N_2(k)$ lie before~$\gamma_k$ and start before~$k$, and the spans in~$N_3(k)$ lie after~$\gamma_k$ and end after~$k$.
    The spans in~$N_4(k)$ avoid~$\gamma_k$ and lie within the span of~$k$, so~$N_4(k)$ is empty if~$k$ is a root.
    The jobs in~$N_0(k)$ are the~$m$ last neighbors of~$k$ in~$\prec$.}
    \label{fig:decomposition}
\end{figure}
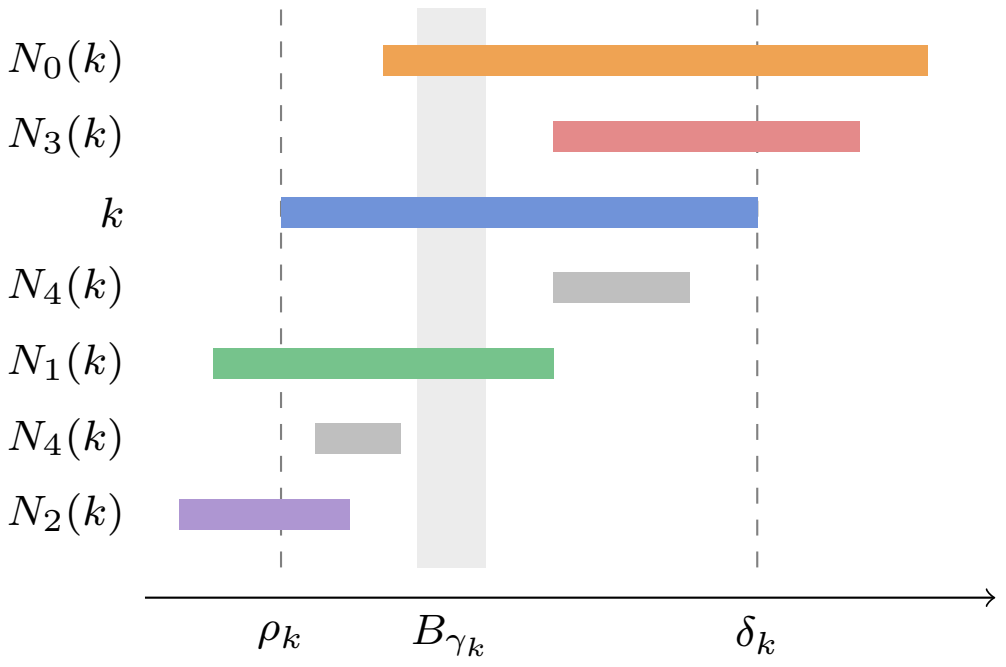

\begin{lemma}
    \label{lem:decomposition}
    The sets $N_0(k),\dots,N_4(k)$ partition~$N(k)$.
\end{lemma}
\begin{proof}
    The sets are pairwise disjoint, since only spans in~$N_1(k)$ contain~$\gamma_k$, spans in~$N_2(k)$ and~$N_3(k)$ are not contained in~$[\rho_k,\delta_k]$, and $\delta_j<\gamma_k<\rho_j$ is impossible.
    Let $j\in N(k)\setminus(N_0(k)\cup N_1(k))$.
    Then $\delta_j<\gamma_k$ or $\gamma_k<\rho_j$.
    In the first case, we have $j\in N_2(k)$ if $\rho_j<\rho_k$, and otherwise $[\rho_j,\delta_j]\subseteq[\rho_k,\delta_k]$ since $\delta_j<\gamma_k\le\delta_k$, thus $j\in N_4(k)$.
    In the second case, we have $j\in N_3(k)$ if $\delta_k<\delta_j$, and otherwise $[\rho_j,\delta_j]\subseteq[\rho_k,\delta_k]$ since $\rho_k\le\gamma_k<\rho_j$, so $j\in N_4(k)$.
    Therefore, the sets partition the neighborhood~$N(k)$.
\end{proof}

By definition, $|N_0(k)|=m$.
The structure lemma bounds~$N_1(k)$.

\begin{corollary}
    \label{cor:bound_N1}
    It holds that $|N_1(k)|\le m-2$.
\end{corollary}
\begin{proof}
    Every job $j\in N_1(k)$ satisfies $j\prec g_k$, because $j\notin N_0(k)$ by~\Cref{lem:decomposition}, and $\rho_j\le\gamma_k\le\delta_j$.
    By~\cref{lem:structure}, all jobs in $N_1(k)\cup\{k,g_k\}$ are processed in block~$\gamma_k$.
    Since $k,g_k\notin N_1(k)$ and at most~$m$ jobs are processed in a block, we obtain the bound $|N_1(k)|\le m-2$.
\end{proof}

To exclude~$N_4(k)$, we consider jobs whose span contains no other span.
We say that a job~$j$ \emph{lies inside} a job~$k\neq j$ if $[\rho_j,\delta_j]\subsetneq[\rho_k,\delta_k]$, or $[\rho_j,\delta_j] = [\rho_k,\delta_k]$ and $j\prec k$.
This relation is transitive and irreflexive.
A job~$k$ is a \emph{root} if no job lies inside~$k$.
Since~$J$ is finite, every job is a root or some root lies inside it.
Thus, roots have inclusion-minimal spans, with equal spans resolved by the order~$\prec$.
Choosing a root eliminates~$N_4(k)$; only the neighbors in~$N_2(k)$ and~$N_3(k)$ remain to be bounded.

\begin{lemma}
    \label{lem:root_degree}
    Let~$k$ be a root with $|N(k)|\ge m$.
    Then 
    \begin{align*}
        |N(k)| \leq 2m-2 + |N_2(k)| + |N_3(k)|.
    \end{align*}
\end{lemma}
\begin{proof}
    For $j\in N_4(k)$, we have $[\rho_j,\delta_j]\subsetneq[\rho_k,\delta_k]$, the intervals are not equal because $\gamma_k\in[\rho_k,\delta_k]$ but $\gamma_k\notin[\rho_j,\delta_j]$.
    Thus, $j$ lies inside~$k$, so $N_4(k)=\emptyset$ as~$k$ is a root by assumption.
    By~\cref{lem:decomposition,cor:bound_N1}, we obtain the bound $|N(k)|\le m+(m-2)+|N_2(k)|+|N_3(k)|$.
\end{proof}

In~\cref{app:root-sequence}, we find a root~$k$ with $|N_2(k)|\le m-1$ and $N_3(k) = \emptyset$, which yields the desired bound $|N(k)|\le 3m-3$.

\subsection{Root Sequence}
\label{app:root-sequence}

\begin{figure}[t]
    \centering
    \begin{tikzpicture}[x=0.45cm,y=0.5cm,every node/.style={font=\scriptsize}]
        \fill[gray!15] (3,-0.2) rectangle (4,4.2);
        \fill[gray!15] (8,-0.2) rectangle (9,4.2);
        \fill[myblue!70] (1,0.3) rectangle (6,0.7);
        \fill[mygreen!70] (2.5,1.3) rectangle (7.5,1.7);
        \fill[myblue!40] (6,2.3) rectangle (10,2.7);
        \fill[myred!60] (5,3.3) rectangle (11,3.7);
        \draw[->] (0,-0.6) -- (12,-0.6);
        \node at (3.5,-1.1) {$B_{\gamma_k}$};
        \node at (8.5,-1.1) {$B_{\gamma_{k'}}$};
        \foreach \y/\lab in {0/k,1/j,2/k',3/e_k}
            \node[left] at (0,\y+0.5) {$\lab$};
    \end{tikzpicture}
    \caption{Schematic illustration of successor in root sequence and~\cref{lem:successor}.
    The first job~$e_k$ of~$N_3(k)$ starts after the critical block~$\gamma_k$ and ends after~$k$, and the successor~$k'$ is a root inside~$e_k$.
    The job~$j\in N_2(k')$ ends before~$\gamma_{k'}$, but its span contains~$\gamma_k$, so $j\in N_1(k)$.}
    \label{fig:successor}
\end{figure}
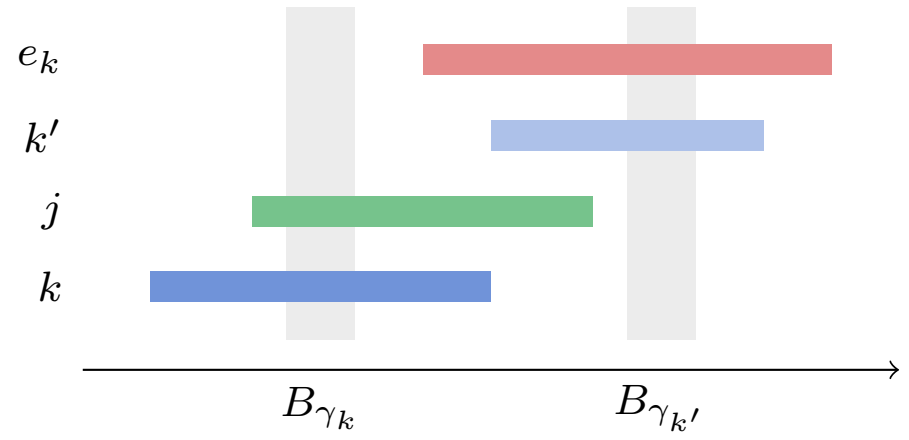

Let~$k$ be a root with $|N(k)|\geq m$ and $N_3(k)\neq  \emptyset$, and let~$e_k$ be the first job of~$N_3(k)$ in~$\prec$.
We define the \emph{successor}~$k'$ of~$k$ as~$e_k$ if~$e_k$ is a root, and otherwise as the first root in~$\prec$ that lies inside~$e_k$.
Thus, the span of~$k'$ is contained in the span of~$e_k$, and no root that lies inside~$e_k$ precedes~$k'$ in~$\prec$.
Moreover, we have that $\rho_k\leq \gamma_k < \rho_{e_k} \leq\rho_{k'}$, since $e_k\in N_3(k)$ (see~\Cref{fig:successor}).
Moving to a successor advances the first processing block.
The next lemma shows that, unless the successor already has degree less than~$m$, its set~$N_2$ has size at most~$m-1$.
Repeating this step eventually reaches a root with small degree or with~$N_3$ empty, completing the degree bound.

\begin{lemma}
    \label{lem:successor}
    Let~$k'$ be the successor of a root~$k$ with $|N(k')|\geq m$.
    Then 
    \begin{align*}
        N_2(k')\subseteq N_1(k) \cup \{k\}.
    \end{align*}
\end{lemma}
\begin{proof}
    Let $e\coloneqq e_k$, and let $j\in N_2(k')\setminus\{k\}$.
    Since $e\in N_3(k)$, we have $\gamma_k<\rho_e$ and $\delta_k<\delta_e$.
    Moreover, $j\prec e$, since $\delta_j<\gamma_{k'}\le\delta_{k'}\le\delta_e$.
    For any~$j'\in N_0(k)$, we have~$e\prec j'$ because $e\in N(k)\setminus N_0(k)$, hence by transitivity~$j\prec j'$, and therefore we have $j\notin N_0(k)$.
    Finally, $\delta_j\ge\rho_{k'} \geq \rho_{e} > \gamma_k$, since~$j$ and~$k'$ are adjacent.
    It remains to show that $\rho_j\le\gamma_k$ and $j\in N(k)$, as then $j\in N_1(k)$ by definition.

    First, we show that $\rho_j\le\delta_k$.
    Otherwise, $\rho_e\le\delta_k<\rho_j$, since~$e$ and~$k$ are adjacent, and $\delta_j<\delta_e$.
    Hence, the span of~$j$ lies strictly inside the span of~$e$.
    Let~$u$ be~$j$ or a root that lies inside~$j$.
    Then~$u$ is a root that lies inside~$e$ with $\delta_u\le\delta_j<\delta_{k'}$, so $u\prec k'$, contradicting the choice of~$k'$.

    Next, we show that $\rho_j\le\gamma_k$.
    Suppose that $\gamma_k<\rho_j\le\delta_k$.
    Since $\rho_k\le\gamma_k<\rho_j$ and~$k$ is a root, $j$ does not lie inside~$k$, so $\delta_k<\delta_j$ and $k\prec j$.
    If $j\notin N(k)$, then we have $\rho_j<\delta_k$, $k\notin\cS_{\rho_j}$, and $j\notin\cS_{\delta_k}$, so a priority swap exists for $k\prec j$ at blocks $\rho_j<\delta_k$.
    Thus, $j\in N(k)\setminus N_0(k)$ with $\gamma_k<\rho_j$ and $\delta_k<\delta_j$, so $j\in N_3(k)$.
    Since $j\prec e$, this contradicts the choice of~$e$.

    Finally, we show that $j\in N(k)$.
    If $\delta_j\le\delta_k$ and $k\prec j$, then $\delta_j=\delta_k$, so both~$j$ and~$k$ are processed in block~$\delta_k$.
    If $\delta_j\le\delta_k$ and $j\prec k$, then $j\in\cS_{\gamma_k}$ or $k\in\cS_{\delta_j}$, as otherwise a priority swap exists for $j\prec k$ at blocks $\gamma_k<\delta_j$.
    In both cases, we have $j\in N(k)$, because $k\in\cS_{\gamma_k}$.
    If $\delta_k<\delta_j$, then $\rho_j\le\gamma_k\le L_{j'k}\le\delta_k<\delta_j$ for every $j'\in N_0(k)$.
    If $j\notin\cS_{L_{j'k}}$, then $j'\in\cS_{\delta_j}$, as otherwise a priority swap exists for $j\prec j'$ at blocks $L_{j'k}<\delta_j$.
    Since~$\cS_{\delta_j}$ contains~$j$ and at most $m-1$ other jobs, we obtain $j\in\cS_{L_{j'k}}$ for some $j'\in N_0(k)$, and thus $j\in N(k)$.
\end{proof}

\begin{lemma}
    \label{lem:low_degree}
    Some job has at most $3m-3$ neighbors.
\end{lemma}
\begin{proof}
    Starting from the first root in~$\prec$, we repeatedly move to the successor as long as it exists.
    Since $\rho_k\leq \gamma_k < \rho_{k'}$ for every root~$k$ with successor~$k'$, this terminates in a root~$k$ with $|N(k)|<m$ or~$N_3(k)=\emptyset$.
    If $|N(k)|<m$, we are done, so assume $N_3(k) = \emptyset$.
    We show that $|N_2(k)|\leq m-1$.
    If~$k$ is the successor of a root~$k_0$, then $N_2(k) \subseteq N_1(k_0) \cup \{k_0\}$ by~\cref{lem:successor}, so $|N_2(k)| \leq (m-2)+1$ by~\cref{cor:bound_N1}.
    Otherwise, $k$ is the first root in~$\prec$, and $N_2(k) = \emptyset$.
    Indeed, for $j\in N_2(k)$, the job~$j$ or a root that lies inside~$j$ ends no later than $\delta_j < \gamma_k\leq \delta_k$, and is thus a root preceding~$k$.
    We conclude by~\cref{lem:root_degree} that~$|N(k)|\leq 2m-2+(m-1)=3m-3$.
\end{proof}

We now prove the upper bound of~\cref{thm:canonical_3m-2}.
We show tightness of the chromatic bound in~\cref{app:canonical-tightness}.

\begin{proof}
    For every nonempty $J'\subseteq J$, the schedule with processing sets $\cS_t\cap J'$ is canonical, and its conflict graph is $G[J']$.
    Applying~\cref{lem:low_degree} to this schedule yields $\delta(G[J'])\le 3m-3$.
    Hence,~$\chi(G)\le 3m-2$ by~\eqref{eq:szekeres_wilf}.
\end{proof}

\subsection{Tightness}
\label{app:canonical-tightness}

We show that the bound of~\cref{thm:canonical_3m-2} cannot be improved using canonicity alone.

\begin{figure}[t]
    \centering
    \begin{subfigure}[b]{0.62\textwidth}
        \centering
        \begin{tikzpicture}[x=0.5cm,y=0.5cm,every node/.style={font=\scriptsize}]
            \foreach \x in {3,5,8}
                \draw[dashed,gray] (\x,-0.2) -- (\x,7.2);
            \foreach \y in {0,1}
                \fill[myblue!70] (0,\y+0.15) rectangle (5,\y+0.85);
            \fill[myorange!80] (3,2.15) rectangle (4,2.85);
            \fill[myorange!80] (5,2.15) rectangle (8,2.85);
            \fill[myorange!80] (4,3.15) rectangle (8,3.85);
            \fill[mygreen!70] (2,4.15) rectangle (3,4.85);
            \fill[mygreen!70] (5,4.15) rectangle (6,4.85);
            \fill[mygreen!70] (8,4.15) rectangle (9,4.85);
            \fill[mygreen!70] (1,5.15) rectangle (2,5.85);
            \fill[mygreen!70] (6,5.15) rectangle (7,5.85);
            \fill[mygreen!70] (8,5.15) rectangle (9,5.85);
            \fill[mygreen!70] (0,6.15) rectangle (1,6.85);
            \fill[mygreen!70] (7,6.15) rectangle (9,6.85);
            \draw[->] (0,-0.6) -- (9.6,-0.6);
            \node at (1.5,-1.1) {$U\cup\{w_i\}$};
            \node at (4,-1.1) {$U\cup\{v_i\}$};
            \node at (6.5,-1.1) {$V\cup\{w_i\}$};
            \node at (8.5,-1.1) {$W$};
            \foreach \y/\lab in {0/u_1,1/u_2,2/v_1,3/v_2,4/w_1,5/w_2,6/w_3}
                \node[left] at (0,\y+0.5) {$\lab$};
        \end{tikzpicture}
    \end{subfigure}
    \hfill
    \begin{subfigure}[b]{0.34\textwidth}
        \centering
        \begin{tikzpicture}[every node/.style={font=\scriptsize},
            vtx/.style={circle,draw,inner sep=0pt,minimum size=13pt}]
            \foreach \i/\lab/\col in {0/u_1/myblue!70,1/u_2/myblue!70,2/v_1/myorange!80,3/v_2/myorange!80,4/w_1/mygreen!70,5/w_2/mygreen!70,6/w_3/mygreen!70}
                \node[vtx,fill=\col] (n\i) at ({90-360*\i/7}:1.25cm) {$\lab$};
            \foreach \i [evaluate=\i as \s using int(\i+1)] in {0,...,5}
                \foreach \j in {\s,...,6}
                    \draw (n\i) -- (n\j);
        \end{tikzpicture}
    \end{subfigure}
    \caption{The canonical schedule of~\cref{prop:canonical_lower_bound} for $m=3$, with unit blocks.
    Left:~The dashed lines separate the four phases, and the labels below indicate their processing sets.
    Right:~Every pair of jobs is processed together, so the conflict graph is the complete graph~$K_7$.}
    \label{fig:canonical_lower_bound}
\end{figure}
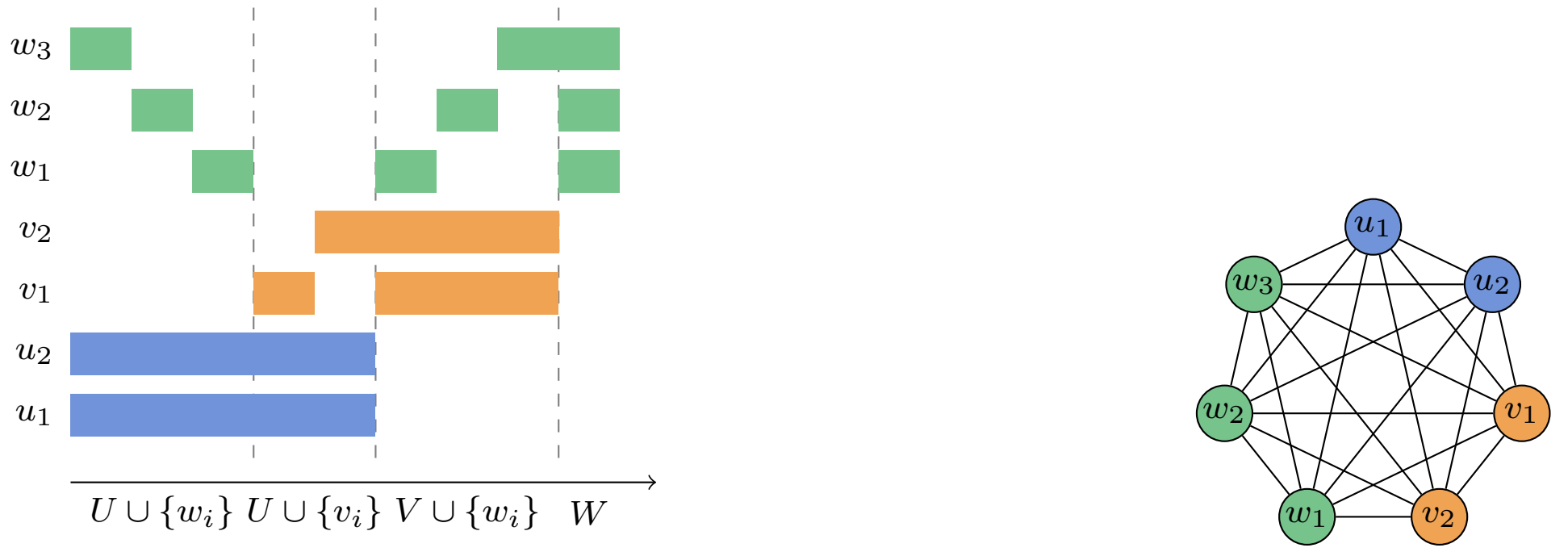

\begin{proposition}
    \label{prop:canonical_lower_bound}
    For every integer $m\ge2$, there is a canonical schedule~$\cS$ on~$m$ machines with the conflict graph $G^\cS$ being a clique on~$3m-2$ jobs.
\end{proposition}
\begin{proof}
    Let $U\coloneqq \{u_1,\dots,u_{m-1}\}$, $V\coloneqq\{v_1,\dots,v_{m-1}\}$, and $W\coloneqq \{w_1,\dots,w_m\}$, and let $J\coloneqq U\cup V\cup W$, so $|J|=3m-2$.
    We index the jobs in the order $u_1,\dots,u_{m-1},v_1,\dots,v_{m-1},w_1,\dots,w_m$.
    The schedule~$\cS$ has the unit blocks $B_t=[t-1,t)$ for $t\in[3m]$ with processing sets
    \begin{align*}
        \cS_t \coloneqq
        \begin{cases}
            U\cup\{w_{m+1-t}\} &\text{if } t\in \{1,\dots,m\}, \\
            U\cup\{v_{t-m}\} &\text{if } t\in \{m+1,\dots,2m-1\}, \\
            V\cup\{w_{t-2m+1}\} &\text{if } t\in \{2m,\dots,3m-1\}, \\
            W &\text{if } t=3m.
        \end{cases}
    \end{align*}
    That is, the jobs of~$U$ run together with $w_m,\dots,w_1$ and then with $v_1,\dots,v_{m-1}$.
    Afterwards, the jobs of~$V$ run together with $w_1,\dots,w_m$, and finally the jobs of~$W$ run together (see~\Cref{fig:canonical_lower_bound}).
    Every processing set contains exactly~$m$ jobs.
    For every job~$j$, define the processing time~$p_j$ as the number of blocks in which~$j$ is processed, furthermore, define $r_j\coloneqq b_{\rho_j}$ and $d_j\coloneqq b_{\delta_j+1}$.
    Hence, $\cS$ is a feasible schedule on~$m$ machines.

    The jobs of~$U$, $V$, and~$W$ have their last blocks $2m-1$, $3m-1$, and~$3m$, respectively.
    With ties broken by index, the canonical ordering~$\prec$ is the order in which we indexed the jobs.

    We show that~$\cS$ admits no priority swap.
    Suppose that a priority swap exists for $j_1\prec j_2$ at blocks $t_1<t_2$.
    Since $\rho_{j_1}\le t_1$, $j_1\notin\cS_{t_1}$, and $j_1\in\cS_{t_2}$, we have $\rho_{j_1}<t_1<t_2\le\delta_{j_1}$.
    Thus, $j_1$ pauses in block~$t_1$.
    The jobs of~$U$ are processed in all blocks $1,\dots,2m-1$ and never pause, so $j_1\notin U$.

    Suppose that $j_1=v_i$.
    Job~$v_i$ is processed in block~$m+i$ and in all blocks $2m,\dots,3m-1$.
    Hence, $t_1=m+l$ for some $l>i$.
    Since $j_2\in\cS_{m+l}=U\cup\{v_l\}$ and $v_i\prec j_2$, we have $j_2=v_l$.
    However, $v_l$ is processed in all blocks $2m,\dots,3m-1$, and in particular in block~$t_2$, contradicting $j_2\notin\cS_{t_2}$.

    Suppose that $j_1=w_i$.
    Since $w_i\prec j_2$, we have $j_2=w_l$ for some $l>i$.
    Job~$w_l$ is processed in the blocks $m+1-l$, $2m-1+l$, and~$3m$.
    We have $m+1-l<m+1-i=\rho_{w_i}<t_1$, and $w_i\in\cS_{3m}$, so $t_1=2m-1+l$.
    Since $2m-1+i<t_1$, the only block after~$t_1$ in which~$w_i$ is processed is~$3m$.
    Hence, we have $t_2=3m$, but $w_l\in\cS_{3m}$, contradicting $j_2\notin\cS_{t_2}$.
    Therefore, $\cS$ is canonical.

    Finally, every pair of distinct jobs is processed together in some block.
    Pairs within~$U$, and pairs~$\{u,w_i\}$ with $u\in U$, are processed together in block~$m+1-i$.
    Pairs $\{u,v_i\}$ with $u\in U$ are processed together in block~$m+i$.
    Pairs within~$V$, and pairs $\{v,w_i\}$ with $v\in V$, are processed together in block~$2m-1+i$.
    Pairs within~$W$ are processed together in block~$3m$.
    Thus, we conclude that the conflict graph of the canonical schedule~$\cS$ is a complete graph on~$3m-2$ vertices.
\end{proof}

The schedule of~\cref{prop:canonical_lower_bound} is not supercanonical.
For instance, a delay swap exists for~$w_m$ and~$w_1$ at blocks $\rho_{w_1}=m<3m-1$, since it holds that $\rho_{w_m}=1<m<3m-1<3m=\delta_{w_1}$, $w_m\in\cS_{3m-1}\setminus\cS_m$, and $w_1\in\cS_m\setminus\cS_{3m-1}$.
Moreover, the instance itself admits a non-migratory schedule on $2m-1$ machines by~\cref{thm:main}, so only this particular schedule requires $3m-2$ machines.
This shows that excluding delay swaps is essential for~\cref{thm:coloring}.

\section{Minimum Degree of Supercanonical Conflict Graphs}
\label[appendix]{appendix:supercanonical_min_degree}

We show that a supercanonical conflict graph need not have a vertex of degree at most $2m-2$.
Thus, the Szekeres--Wilf bound~\cite{SzekeresW68} does not suffice to establish $(2m-1)$-colorability.
This motivates the recoloring argument using Kempe exchanges.

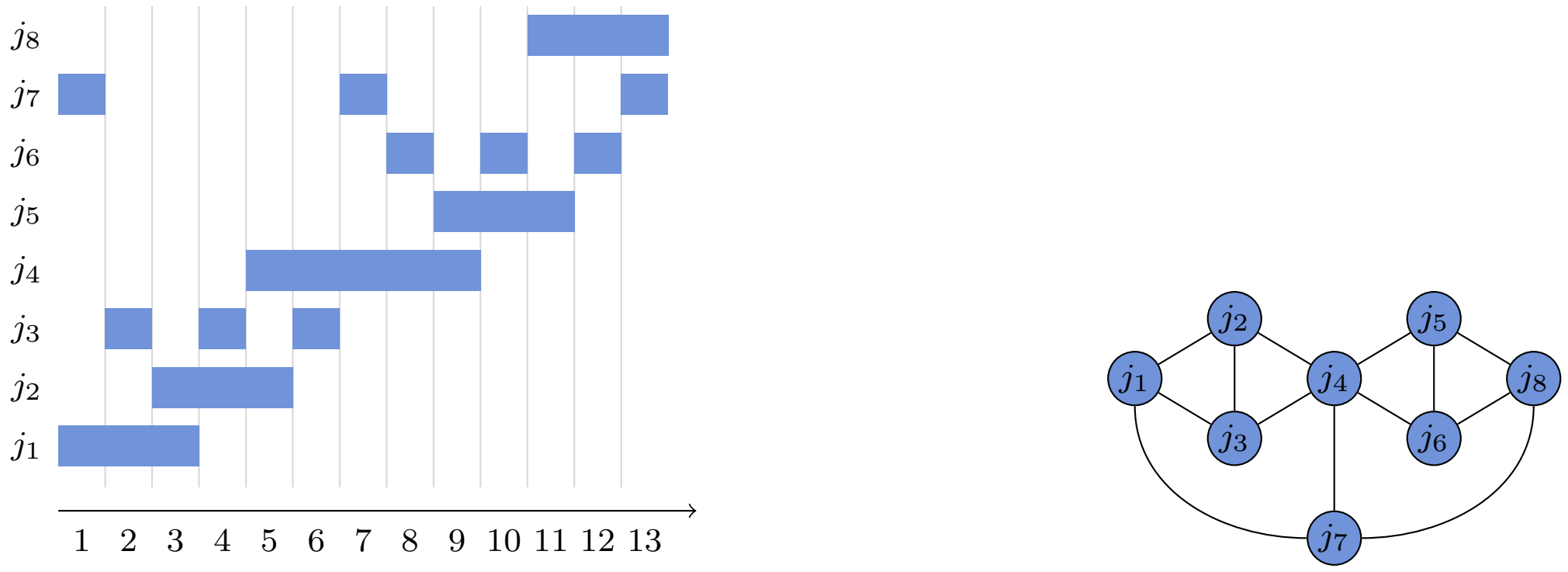
\begin{figure}[t]
    \centering
    \begin{subfigure}[b]{0.6\textwidth}
        \centering
        \begin{tikzpicture}[x=0.4cm,y=0.5cm,every node/.style={font=\scriptsize}]
            \foreach \x in {1,...,12}
                \draw[gray!30] (\x,-0.2) -- (\x,8);
            \foreach \y/\a/\b in {0/0/3, 1/2/5, 2/1/2, 2/3/4, 2/5/6, 3/4/9,
                                  4/8/11, 5/7/8, 5/9/10, 5/11/12,
                                  6/0/1, 6/6/7, 6/12/13, 7/10/13}
                \fill[myblue!70] (\a,\y+0.15) rectangle (\b,\y+0.85);
            \draw[->] (0,-0.6) -- (13.6,-0.6);
            \foreach \t in {1,...,13}
                \node at (\t-0.5,-1.1) {$\t$};
            \foreach \y in {1,...,8}
                \node[left] at (0,\y-0.5) {$j_{\y}$};
        \end{tikzpicture}
    \end{subfigure}
    \hfill
    \begin{subfigure}[b]{0.38\textwidth}
        \centering
        \begin{tikzpicture}[x=0.85cm,y=0.85cm,every node/.style={font=\scriptsize},
            vtx/.style={circle,draw,fill=myblue!70,inner sep=0pt,minimum size=13pt}]
            \node[vtx] (j1) at (0,1)   {$j_1$};
            \node[vtx] (j2) at (1,1.6) {$j_2$};
            \node[vtx] (j3) at (1,0.4) {$j_3$};
            \node[vtx] (j4) at (2,1)   {$j_4$};
            \node[vtx] (j5) at (3,1.6) {$j_5$};
            \node[vtx] (j6) at (3,0.4) {$j_6$};
            \node[vtx] (j7) at (2,-0.6) {$j_7$};
            \node[vtx] (j8) at (4,1)   {$j_8$};
            \draw (j1)--(j2) (j1)--(j3) (j2)--(j3) (j2)--(j4) (j3)--(j4)
                  (j4)--(j5) (j4)--(j6) (j5)--(j6) (j5)--(j8) (j6)--(j8)
                  (j4)--(j7);
            \draw (j1) to[out=-90,in=180] (j7);
            \draw (j8) to[out=-90,in=0] (j7);
        \end{tikzpicture}
    \end{subfigure}
    \caption{The supercanonical schedule of~\cref{prop:supercanonical_min_degree} on two machines, with unit blocks numbered below the axis.
    Left:~Every block contains exactly two jobs, and only~$j_3$, $j_6$, and~$j_7$ pause.
    Right:~In the conflict graph, $j_4$ has degree~$5$ and every other job has degree~$3$, so the minimum degree exceeds $2m-2=2$.}
    \label{fig:supercanonical_min_degree}
\end{figure}

\begin{proposition}
\label{prop:supercanonical_min_degree}
    There is a supercanonical schedule $\cS$ on two machines with minimum degree $\delta(G^\cS)=3$.
\end{proposition}
\begin{proof}
    Let $J\coloneqq \{j_1,\ldots,j_8\}$.
    The schedule $\cS$ has unit blocks $B_t=[t-1,t)$ for $t\in[13]$, with processing sets
    \[
    \begin{aligned}
        \cS_1&=\{j_1,j_7\},&
        \cS_2&=\{j_1,j_3\},&
        \cS_3&=\{j_1,j_2\},\\
        \cS_4&=\{j_2,j_3\},&
        \cS_5&=\{j_2,j_4\},&
        \cS_6&=\{j_3,j_4\},\\
        \cS_7&=\{j_4,j_7\},&
        \cS_8&=\{j_4,j_6\},&
        \cS_9&=\{j_4,j_5\},\\
        \cS_{10}&=\{j_5,j_6\},&
        \cS_{11}&=\{j_5,j_8\},&
        \cS_{12}&=\{j_6,j_8\},\\
        \cS_{13}&=\{j_7,j_8\}.
    \end{aligned}
    \]
    For every job $j$, define $p_j$ as the number of blocks in which $j$ is processed, and let $r_j\coloneqq b_{\rho_j}$ and~$d_j\coloneqq b_{\delta_j+1}$.
    Every processing set contains exactly two jobs, so~$\cS$ is feasible on two machines.
    The last processing blocks are
    \[
        (\delta_{j_1},\ldots,\delta_{j_8})
        =(3,5,6,9,11,12,13,13),
    \]
    so, with ties broken by index, the canonical ordering is $j_1\prec\cdots\prec j_8$.

    Only $j_3,j_6,j_7$ pause between their first and last processing blocks, so only these can be the $\prec$-earlier job of a priority swap.
    During a pause of $j_3$, the only $\prec$-later job processed is $j_4$ in block~$5$.
    However, $j_4$ also runs in block~$6$, the only subsequent processing block of $j_3$.
    Similarly, for $j_6$, the only possibility is $j_8$ in block~$11$, which also runs in block~$12$.
    Finally, the only job later than $j_7$ is $j_8$, which runs in blocks $11,12,13$.
    Its first two blocks precede the final block of $j_7$, which also contains $j_8$.
    Hence, no priority swap exists.

    The later-starting job of a delay swap must also pause.
    Job~$j_7$ starts in the first block, so it cannot be this job.
    During the pauses of $j_3$, the only processed job that starts before $j_3$ is $j_1$, which also runs in the first block of $j_3$.
    Similarly, during the pauses of $j_6$, the only such job is $j_4$, which also runs in the first block of $j_6$.
    Thus, no delay swap exists, and $\cS$ is supercanonical.

    Every job other than $j_4$ has three neighbors, while $j_4$ has five.
    Hence, the minimum degree satisfies $\delta(G^\cS) = 3 > 2 = 2m-2$.
\end{proof}

\end{document}